\documentclass[aps,prd,english,superscriptaddress,11pt,notitlepage]{revtex4}
	\usepackage[colorlinks=true, a4paper=true, pdfstartview=FitV,
linkcolor=blue, citecolor=blue, urlcolor=blue]{hyperref}

\usepackage{amsmath}
\usepackage{amssymb}
\usepackage{graphicx}
\usepackage{hhline}
\usepackage{textcomp}
\makeatletter
\newcommand{\bea}{\begin{eqnarray}}
\newcommand{\eea}{\end{eqnarray}}

\newcommand{\be}{\begin{equation}}
\newcommand{\ee}{\end{equation}}
\usepackage[active]{srcltx}

\begin{document}

 \title{Kink-antikink collisions in a weakly interacting $\phi^4$ model. }

\author{C. Adam}
\email[]{adam@fpaxp1.usc.es}
\affiliation{Departamento de F\'isica de Part\'iculas, Universidad de Santiago de Compostela and Instituto Galego de F\'isica de Altas Enerxias (IGFAE) E-15782 Santiago de Compostela, Spain}
\author{K. Oles}
\email[]{katarzyna.slawinska@uj.edu.pl}
\affiliation{Institute of Physics,  Jagiellonian University, Lojasiewicza 11, Krak\'{o}w, Poland}
\author{T. Romanczukiewicz}
\email[]{tomasz.romanczukiewicz@uj.edu.pl}
\affiliation{Institute of Physics,  Jagiellonian University, Lojasiewicza 11, Krak\'{o}w, Poland}
\author{A. Wereszczynski}
\email[]{andrzej.wereszczynski@uj.edu.pl}
\affiliation{Institute of Physics,  Jagiellonian University, Lojasiewicza 11, Krak\'{o}w, Poland}

\begin{abstract}
Kink-antikink scattering in non-integrable field theories like $\phi^4$ theory is still rather poorly understood beyond brute-force numerical calculations, even after several decades of investigation. 
Recently, however, some progress has been made based on the introduction of certain "self-dual" background fields in these field theories which imply both the existence of static kink-antikink solutions of the Bogomolnyi type and the possibility of an adiabatic scattering ("moduli space approximation"). 

Here we continue and generalize these investigations by introducing a one-parameter family of models interpolating between the BPS (=Bogomolnyi-Prasad-Sommerfield") model with the self-dual background field and the original $\phi^4$ theory. More concretely, we study kink-antikink scattering in a parameter range  between the limit of no static force (BPS limit) and the regime where the static interaction between kink and antikink is small (non-BPS). This allows us to study the impact of the strength of the intersoliton static force on the soliton dynamics.

In particular, we analyze how the transition of a bound mode through the mass threshold  affects the soliton dynamics in a generic process, i.e., when  a static intersoliton force shows up. We show that the thin, precisely localized spectral wall which forms in the limit of no static force, broadens in a well-defined manner when a static force is included,  giving rise to what we will call a thick spectral wall. 
This phenomenon results from the appearance of a stationary saddle point solution where
the acceleration of the solitons owing to the attractive force is compensated by
the dynamics of the sufficiently excited mode.
Thus, now this barrier shows up before the mode crosses the mass threshold.

\end{abstract}

\maketitle

 \vspace*{0.2cm}


\section{Motivation}
Scattering of topological solitons is a complicated process, revealing many nontrivial phenomena even in (1+1) dimensional theories \cite{SM, Shnir, Kev}. Typically, solitons interact in three different ways, i.e., by a static force \cite{SM, Manton:2018deu}, with normal (and even quasi-normal) modes \cite{Sug, Dorey:2017dsn, Dorey:2011yw} which may store and release portions of the energy, and with radiation \cite{Forgacs:2008az, Romanczukiewicz:2017hdu}. During a soliton-antisoliton (SAS) collision, generally, all these types of interactions have a nontrivial impact on the dynamics, which leads to a rather complex pattern of behavior. As a consequence, many phenomena in soliton collisions still defy a conceptual or qualitative understanding. A prototypical example is provided by $\phi^4$ theory where, after 40 years of struggle, there is basically no satisfactory explanation of the observed phenomena occurring in SAS scattering like, e.g., the fractal structure \cite{Sug}-\cite{PLTC}. The underlying problem, however, is not restricted to this example but  concerns a whole variety of solitonic models and processes, even in (1+1) dimensions, especially if the solitons possess internal modes. 

One standard approach is to simplify the dynamics of solitons to a finite dimensional subspace of the full configuration space, i.e., a system of collective coordinates (effective degrees of freedom), usually called moduli. 
They form a space with a metric naturally inherited from the action called moduli space \cite{MORW}.
Then, the reduced dynamics on moduli space should reproduce the most relevant aspects of the full field theoretical dynamics. This strategy works especially well in the case of solitons which do not have static interactions, the so-called Bogomolnyi-Prasad-Sommerfield (BPS) solitons like, for example, the Abelian Higgs vortices \cite{JT}, \cite{Stuart}, \cite{Manton:2002wb} at critical coupling or Yang-Mills-Higgs monopoles \cite{Ma8}. 

BPS solitons share the following useful and interrelated properties.
\begin{itemize}
\item[]{\em i)} The configuration space of the field theory allows to define a topological degree or charge, and the static energy functional has a lower bound which is linear in this topological charge. 

\item[]{\em ii)} For each possible value of the topological charge there exists a whole family of static "BPS" solutions which saturate this lower bound.

\item[]{\em iii)} In addition to the second order static Euler-Lagrange (EL) equations, these BPS solutions fulfill certain first-order equations usually referred to as Bogomolnyi or self-duality equations. 

\item[]{\em iv)} The Bogomolnyi equations have more symmetries than the original EL equations, and the parameters describing this enhanced symmetry give rise to the full family of BPS solutions and provide a natural moduli space. 

\item[]{\em v)} When the spectrum of linear fluctuations about a given soliton is analyzed, then these symmetries or flat directions induce "vibrational" modes with zero frequency, also known as zero modes. 
\end{itemize}

In general, the moduli space for such systems defines a riemannian manifold whose coordinates describe a space of static $N$-soliton configurations with the same (potential) energy, usually related to the arbitrary positions (and possible orientations) of the individual solitons. In this case, the lowest order dynamics is just a slow transition between different solutions with the same static energy, which is induced by the excitations of the respective zero modes and fully controlled by the metric on the moduli space, i.e., by the kinetic energy for low velocities. This approach can be generalized to solitons being close to BPS, i.e., where the static inter-soliton force is small. Then, the moduli space dynamics is governed not only by the metric on the moduli space but also by the effective potential. This is the unstable manifold construction \cite{Manton-1}, which successfully reproduces scattering of weakly interacting solitons, see, e.g., \cite{Speight:1996px} (for an application to the quantum regime see \cite{Halcrow:2020lez}).

There is, however, no canonical moduli space for SAS collisions. The naive use of the unstable manifold leads to some difficulties \cite{weigel-2}. One reason is that, in an SAS process, at some point the solitons interact strongly. This is a violent process, especially when they are close to a vacuum state. Another obstacle in the application of the unstable manifold is that in an SAS collision the final state, i.e., the vacuum, has much lower energy than the initial state (the infinitely separated SAS pair). Therefore, the annihilation occurs through energetically very distinct configurations which cannot be covered by the naive moduli space procedure, where energy is conserved. Indeed, the naive construction of the moduli space applied to SAS scattering in $\phi^4$ theory gives completely wrong results \cite{weigel-2}. In this specific case, the chosen collective coordinates were the positions of the solitons, $\pm a$, and the shape mode amplitudes $A,B$ excited on each of the solitons. An improved construction of the moduli space is, thus, one important line of investigation which has recently allowed to solve the long standing null vector problem \cite{MORW} (see also \cite{PLTC} for some new results). 

Another important source of problems which has not been taken into account by moduli space approximations is that any bound mode, which exists on a free soliton, must necessarily disappear during the annihilation. In non-annihilating collisions (e.g., soliton-soliton or soliton-localized impurity), the bound modes do not have to necessarily cross the mass threshold, but nevertheless the spectral structure---the frequencies of the modes---may change significantly \cite{GoodmanHaberman}. In SAS collisions all the bound modes must enter the continuous spectrum, because the SAS configurations describing the collision must pass through the vacuum configuration which cannot support any bound modes. This fact was ignored in all effective model approaches, where the full field theory dynamics was replaced by a small number of (the lightest) degrees of freedom \footnote{In soliton-soliton scattering the normal mode structure also varies during the collision. However, this change is less drastic as the number of  modes is often a linear function of the topological charge \cite{Gudnason:2018ysx} and therefore is conserved.}. 

Recently, a novel approach based on the so-called self-dual background field framework has been proposed \cite{imp, imp-phi4, susy-imp, solvable-imp}, in which the changeable nature of the bound modes and their impact on the soliton dynamics can be treated in a systematic and well-defined way. The unique advantage of this approach is that it allows to disentangle the role played by internal modes in the soliton dynamics and, therefore, it nontrivially enlarges the insight provided by the standard moduli space approximation. Specifically, for a given solitonic process in a given (1+1) dimensional field theory $L[\phi]$, there always exists a background field $\sigma$ (an impurity) which transforms this process into a self-dual one (i.e., a process where there are no static interactions between the constituent solitons), by a certain deformation of the initial Lagrangian
\be
 L[\phi] \rightarrow L[\phi, \sigma].
 \ee
Physically, e.g., in the case of SAS scattering, this means that there is no static force between the colliding soliton and antisoliton \cite{solvable-imp}, and they can be placed at any distance from each other. Thus, there is a whole family of infinitely many static SAS solutions with the same energy, giving rise to the appearance of a natural, {\it canonical} moduli space.  Furthermore, as the solitons change their position (which corresponds to a flow on the moduli space), their spectral structure also changes. Therefore, the impact of the modes (and their changing nature) on the soliton dynamics can be disentangled. The main conclusion is that in this zero-static-force limit the dynamics of solitons is governed by two new phenomena. Namely, by {\it spectral walls} (SW) and their higher-order counterparts, as well as by the {\it vacuum wall} (VW). We will briefly explain these concepts in the next section.

As said already, in a generic SAS scattering in (1+1) dimensional scalar field theories---like, for example, in the $\phi^4$ model---static soliton interactions are, of course, {\em not} turned off. This means that, even in the limit of very slow velocities, the collision does not happen via a sequence of energetically equivalent (BPS or self-dual) states. In other words, solitons act with a static force on each other which influences the outcome and complexity of the scattering process in a nontrivial manner. Thus the natural question arises how the results relevant for the zero-static-force limit of the SAS collision are modified if a non-zero force is re-introduced. This is related to the second part of our framework in which the self-dual background field should be switched off
\be
 L[\phi, \sigma] \rightarrow L[\phi].
 \ee
This should be performed in a controllable, perturbative way by a smooth change of a deformation parameter $\epsilon$, which provides an interpolation between the zero force limit $L[\phi, \sigma]$ and the original theory $L[\phi]$, 
\be
L[\phi, \sigma; \epsilon=0] \equiv L[\phi, \sigma] \rightarrow L[\phi] =L[\phi, \sigma; \epsilon=1].
\ee

In the present work we make the first step in this project and consider a version of $\phi^4$ theory where kink and antikink interact in a very weak manner, which corresponds to the regime where $\epsilon \ll 1$. The main aim is to understand the fate of the structures found in the zero-force limit. Specifically, we want to verify whether phenomena like spectral walls and/or a vacuum wall continue to exist also for generic solitonic collisions, where static kinks do interact. An affirmative answer definitely matters for their relevance for realistic processes.

As an example, we choose the $\phi^4$ theory as a representative model with topological kinks. We want to emphasize, however, that similar results can be obtained for other (1+1) dimensional solitonic field theories. Let us also underline that the specific model we consider enjoys {\it the robust qualitative features of a generic SAS scattering}, that is, the existence of a non-vanishing static force. However, being a (small) deformation of the (self-dual) theory with vanishing static force, it provides a solid mathematical ground where the resulting phenomena can be studied in a well-defined  and quantitatively reliable way. This allows us to conjecture that the results we obtain are robust and relevant for rather generic soliton-antisoliton scattering processes in (1+1) dimensions. 

The paper is organized as follows. In the next section we present a brief review of the $\phi^4$ model with the self-dual background. In particular, we reveal its SAS BPS solutions and their spectral structure, and we explain the concepts of spectral wall (SW) and vacuum wall (VW). In section 3, we introduce the one-parameter family of models interpolating between the BPS model and the standard $\phi^4$ theory. Further, we explain how the "moduli space approximation" of the BPS model generalizes to the "unstable manifold approximation" for a weak breaking of the BPS property, corresponding to a sufficiently small value of the interpolation parameter $\epsilon$.   In section 4 we study kink-antikink collisions numerically for small $\epsilon$ (weakly broken self-duality). In particular, we investigate how the SW and VW change when a small static intersolitonic force is re-introduced in the model. We find that the VW quickly disappears for growing $\epsilon$, while the SW continues to exist but slightly changes its characteristic properties. In section 5, we study these changes in detail, combining numerical and analytical methods, and introduce and explain the related concept of a thick spectral wall. Section 6 contains our summary.


\section{The $\phi^4$ model without static intersoliton forces}
In this section we briefly summarize those results of \cite{solvable-imp} and \cite{no-force-scatt} which are relevant for our purposes. 
\subsection{The Bogomolny kink-antikink solutions and the moduli space}
Contrary to the standard $\phi^4$ model (and other scalar field theories in one spatial dimension), the self-dual background field deformation allows for static kink-antikink solutions. In fact, for such deformations there exist infinitely many static solutions which saturate a pertinent topological bound and obey a version of the Bogomolny equation. They form a space of energetically equivalent solutions parameterized by a parameter which provides the coordinate on the one dimensional moduli space. Here we briefly summarize the most relevant results obtained for the $\phi^4$ model in the background of such a deformation.

The self-dual background field deformation is defined by the following Lagrangian 
\begin{equation} \label{L-sd}
L[\phi, \sigma]=\int_{-\infty}^\infty  dx \left[\frac{1}{2}\phi_t^2 - \frac{1}{2}
\left( \phi_x  + \sigma W'  \right)^2  \right] 
 \end{equation}
where $W$ is a useful auxiliary function which is related to the scalar field potential $V$ via $(1/2) W'^2 =V$. In particular, for the $\phi^4$ model 
 \begin{equation} \label{superpot}
 W=\phi - \frac{1}{3}\phi^3 \quad \Rightarrow \quad W' = 1-\phi^2,
\end{equation}
and $W' = (d/d\phi)W$. 
For a constant background field, e.g., $\sigma = 1$, we recover the standard $\phi^4$ model with potential $(1/2) W'^2 = (1/2) (1-\phi^2)^2$, up to a total derivative (boundary term).   
Further, to avoid clumsy expressions, we shall use the notations $(\partial/\partial t) \phi =\phi_t = \dot \phi$ and 
 $(\partial / \partial x) \phi = \phi_x = \phi '$ for the time and space derivatives.
 
\begin{figure}
\includegraphics[width=8.1cm]{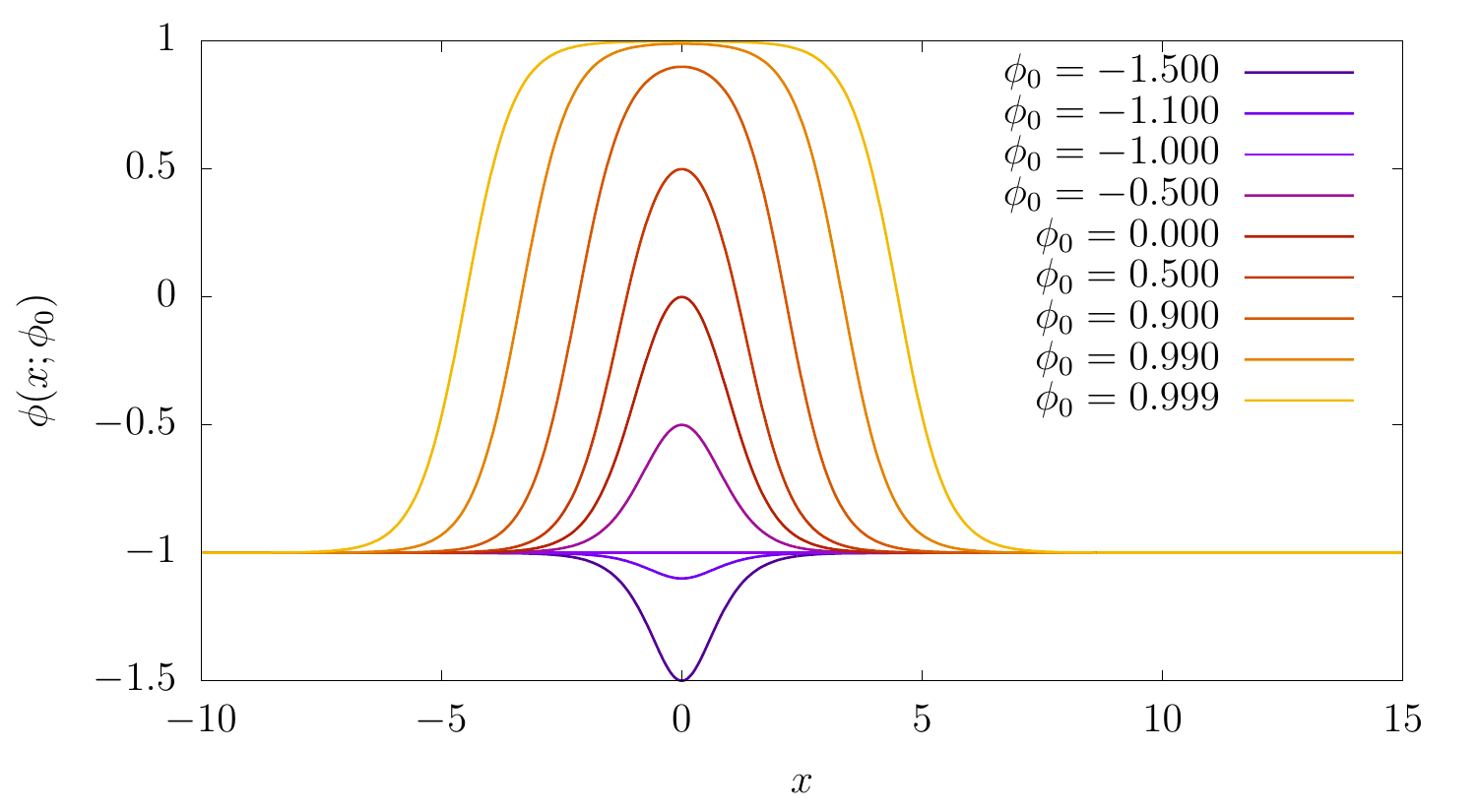}
\includegraphics[width=8.1cm]{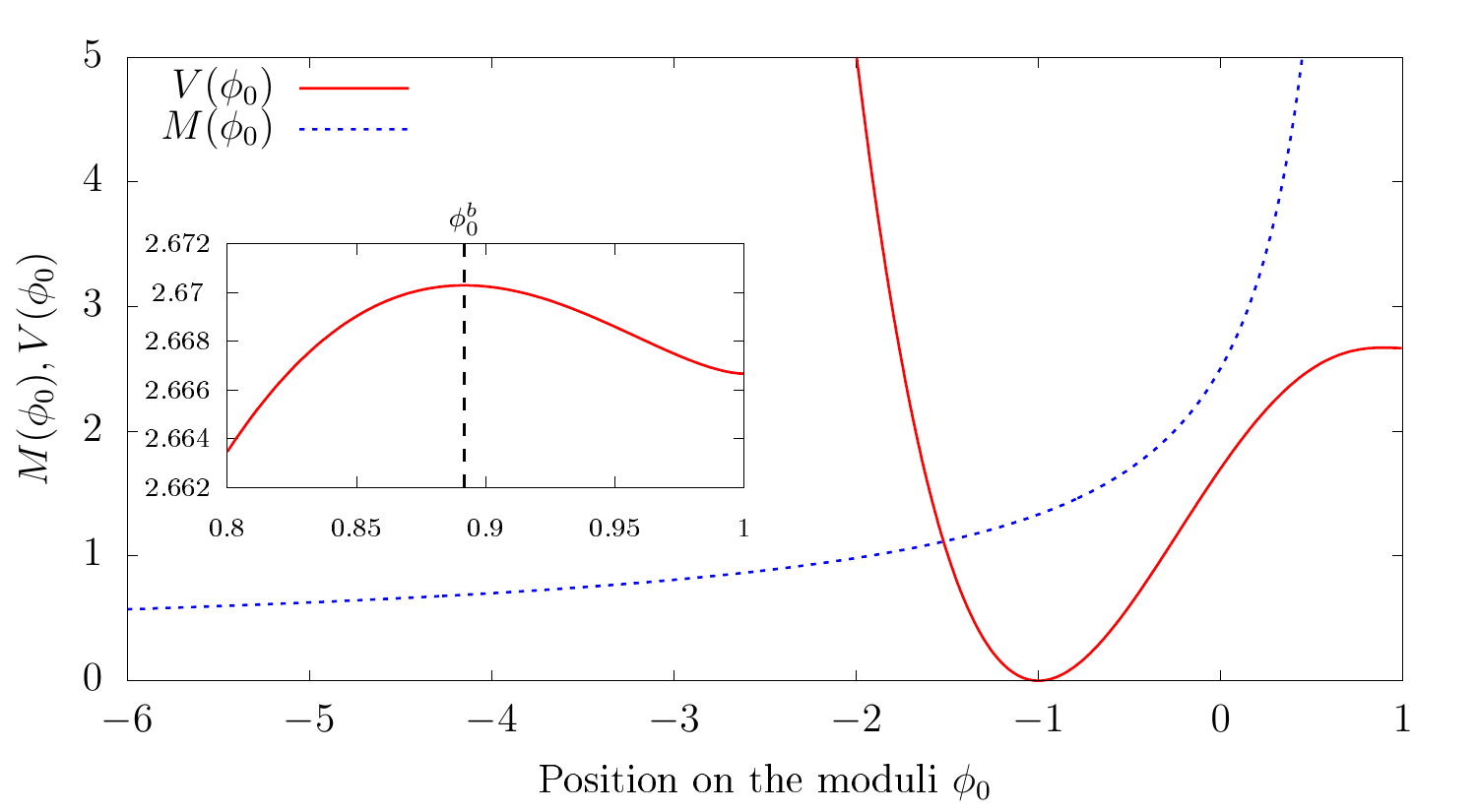}
\caption{Left: SAS solutions (\ref{sol-sd}) for some values of the moduli space coordinate. Right: The moduli space metric $M$ (\ref{mod-metric}) and the effective potential $ V$ (\ref{eff_pot}).
}
\label{effectiv}
\end{figure}

Basically, the background (non-dynamical) field $\sigma$ can have an arbitrary form, 
giving rise to a BPS equation. For the particular case of {\em kink-antikink} BPS solutions in the $\phi^4$ model, however, the relevant background is \cite{solvable-imp}, \cite{no-force-scatt}, \cite{Nick}
 \be \label{imp}
\sigma= \tanh x.
\ee  
Note that $\sigma \to \pm 1$ as $x \to \pm \infty$, which explains the existence of kink and antikink at the same time. 
Then, the static solutions obey the background field deformed Bogomolny equation 
\be
\phi_x=-\tanh x (1-\phi^2)
\ee
forming a one-parameter family of solutions. All solutions of this family have exactly the same (zero) energy and saturate a topological energy bound, therefore the parameter distinguishing different solutions forms a one-dimensional moduli space. Depending on the explicit choice of the moduli space coordinate, the solutions may be expressed, e.g., like
\begin{equation}
\phi (x; a)= \frac{a -\cosh^2 x}{a +\cosh^2 x} \label{sol-sd-a}
\end{equation}
or like
\begin{equation}
\phi(x; \phi_0)= \frac{(1+\phi_0) - (1-\phi_0)\cosh^2 x}{(1+\phi_0) +(1-\phi_0)\cosh^2 x}, \label{sol-sd}
\end{equation}
where the moduli space coordinates $a\in (-1,\infty) $ and $\phi_0\in (-\infty, 1)$ are related via $a\equiv (1+\phi_0)/(1-\phi_0)$.
Here, the expression in terms of $a$ is simpler, but $\phi_0$ has a simpler interpretation. Indeed, it is just the value of the field at the origin, $\phi_0 = \phi (x=0, \phi_0)$. We also remark that both moduli coordinates can be related to the position of the kink and antikink forming the BPS solution, as long as the latter can be defined, see Appendix A. In constrast to $a$ or $\phi_0$, however, the kink (antikink) position is not a good moduli space coordinate, because it does not cover the full moduli space. For $\phi_0 \rightarrow 1$ the solution describes a widely separated pair of a $\phi^4$ kink and antikink. Note that the boundary points $\phi_0=1$ or $a=-1$, respectively, do not belong to the moduli space, which has the topology of the real line. As $\phi_0$ decreases, the solitons approach each other, losing their identity for $\phi_0 <0$.  For $\phi_0=-1$ we get the vacuum solution $\phi=-1$. Then, for $\phi_0 < -1$ the solution reveals a negative bump (below $-1$), which gets deeper for decreasing $\phi_0$ and whose bottom, i.e., $\phi_0$, may approach $-\infty$. This negative bump is, in fact, a welcome property of the moduli space flow, because field values of $\phi (x=0) \equiv \phi_0 <-1$ are reached also in SAS annihilation processes in the pure $\phi^4$ model. Of course, the difference is that $\phi_0 $ cannot take arbitrarily large negative values in the latter case, because a repulsive core forms in the non-self-dual case, see below. Some particular solutions (for different values of the moduli $\phi_0$) are displayed in the left panel of Fig. \ref{effectiv}.

As usual, the moduli space metric $M(\phi_0)$  is calculated by promoting the moduli space coordinate $a$ (or $\phi_0$) to a time-dependent (dynamical) variable, and by inserting it into the self-dual Lagrangian (\ref{L-sd}).  In the self-dual case, only the kinetic term $(1/2)\int dx \dot\phi^2$ contributes, because the static energy is zero for the SAS solutions with trivial topology. With $(1/2)\int dx [(d/dt) \phi(x;\phi_0)]^2 \equiv (1/2)M(\phi_0)\dot\phi_0^2$ we get
(see Fig. \ref{effectiv}, right panel, curve $M(\phi_0)$)
\begin{eqnarray} \label{mod-metric}
M(\phi_0)&=& \int_{-\infty}^\infty dx \left( \frac{d}{d\phi_0}  \phi(x; \phi_0) \right)^2 = \left(\frac{da}{d\phi_0}\right)^2 M(a)\noindent \\ M(a)&=& \frac{1+a}{24a} \left( -3+4a(4+a) +\frac{3(1+6a)}{\sqrt{(1+a)a}} \mbox{arctanh} \sqrt{\frac{a}{1+a}} \; \right) , 
\end{eqnarray}
where the $x$ integration can be done exactly; for compactness, we explicitly show the moduli space metric in the coordinate $a$ in the last expression. One can verify that the boundaries of the moduli space $\phi_0=-\infty$ and $\phi_0=1$ are not attained within a finite time \cite{solvable-imp}. This implies the completeness of the moduli space \cite{MORW}.

As is typical for the BPS models, the simplest dynamics is fully captured by the flow on the moduli space. Here, we consider the initial state
\be
\phi_{in}(x,t)=\tanh(\gamma(x+x_0-vt))-\tanh(\gamma(x-x_0+vt))-1, \label{init-unpert}
\ee
which is a widely separated kink-antikink pair ($x_0\to \infty$) with an initially non-zero kinetic energy, that is, initial velocity $v$. Note that we assume rather non-relativistic velocities for which  $\gamma=1/\sqrt{1-v^2}\approx 1$. For sufficiently large $x_0$, this initial configuration is arbitrarily close to a member of the family of BPS solutions with the identification $\cosh^2 x_0 = 2/(1-\phi_0^{in})$ and $v=-\dot{\phi}^{in}_0/2(1-\phi_0^{in})$, see Appendix B. Then the moduli flow is just a smooth transition between a sequence of  BPS solutions and it is completely described by the metric on the moduli space.

The moduli space approximation, which is the adiabatic motion of the modulus provided by the metric $M$, reproduces very accurately the evolution of the considered initial configuration (\ref{init-unpert}) in the full Euler-Lagrange equation of motion
\begin{equation} \label{BPS-EL-eq}
 \phi_{tt}-\phi_{xx}-\frac{1}{\cosh^2x}(1-\phi^2) - 
2\tanh^2x \, \phi(1-\phi^2)  = 0,
\end{equation}
see Fig. \ref{wall-zero} below. 

{\em Remark: \,} Before continuing, we want to comment on an issue of notation. As mentioned already, the symbol $\phi_0$ in Eq. (\ref{sol-sd}) refers, in fact, to two conceptually different objects whose values, however, coincide. On the one hand, $\phi_0$ is the parameter (moduli space coordinate) which enters at the right hand side of Eq. (\ref{sol-sd}). On the other hand, it is the value which the field $\phi (x,\phi_0)$ takes at the point $x=0$, i.e., $\phi_0 = \phi (x=0, \phi_0)$.  This identification continues to hold in the moduli space approximation, i.e., for a time-dependent $\phi_0 (t)$ in the self-dual case. In the case of a weak breaking of self-duality, (\ref{sol-sd}) is no longer a solution. On the other hand, $\phi_0 (t) \equiv \phi (x=0,t)$ can be defined for arbitrary solutions, by construction. We will continue to use the notation $\phi_0 (t)$ also for the field evaluated at $x=0$, for the following reasons. We will always assume that both the velocities and the self-duality breaking are small ($v<<1$ and $\epsilon <<1$), such that we are in some sense close to the BPS solution. Further, we shall compare the results of the numerical solutions of the full Euler-Lagrage equations to the effective dynamics on the moduli space (more precisely, on the unstable manifold, which is the generalization of the moduli space in the non-selfdual case). For slightly more general initial conditions, we will compare the full dynamics to the effective dynamics on a slightly enlarged space where some normal modes are included in addition to the moduli space. In all cases, the most important source of information is a comparison of the full dynamics of $\phi_0 (t) \equiv \phi (x=0,t)$ with the effective dynamics of the genuine moduli space coordinate $\phi_0 (t)$. For instance, we shall compare the presence of a stationary solution $\phi_0 \sim \;$const. in the full dynamics with the position of a "spectral wall" $\phi_0 =\phi_{sw}=\;$const. in the effective dynamics. Further, we use the relation between the value of $\phi_0$ and the positions of the kink and antikink provided by Eq. (\ref{sol-sd}) also for our interpretation of the full solutions $\phi (x=0,t)$.

\subsection{Flow of the spectral structure}
Because of the fact that each BPS solution is a static solution, we can perform the standard linear perturbation and find the mode structure at each value of the moduli coordinate $\phi_0$. It is a very special and interesting feature of the self-dual background field models that the spectral structure (the spectrum of the discrete modes) flows while we move on the moduli space. Indeed, if we consider a small perturbation of the static solution 
\begin{equation}
\phi(x,t)=\phi(x;\phi_0) + \eta(x,t; \phi_0),
\end{equation}
then, at the linear order in $\eta$, we get the standard linear problem defining the corresponding mode structure
\begin{equation}
\left(-\frac{d^2}{dx^2}+ \frac{2}{\cosh^2x} \phi(x; \phi_0)  - 2\tanh^2 x (1-3\phi^2(x; \phi_0)) \right)\eta=\omega^2(\phi_0)\eta .
\end{equation}
Here we used that $\eta(x,t;\phi_0)=\eta (x; \phi_0) e^{i\omega (\phi_0) t}$ where $\eta (x; \phi_0)$ is the mode and $\omega (\phi_0)$ is its frequency. The dependence on the moduli space parameter $\phi_0$ is explicitly indicated. 
In order to find eigenfrequencies for each value of $\phi_0$,  we have used a
shooting method, starting from even ($\eta(0)=1, \eta'(0)=0$) or odd 
($\eta(0)=0, \eta'(0)=1$) initial conditions. By changing the frequency
we were trying to smoothly match with the asymptotic form $\eta\approx A
e^{-kx}$ at large distance.  The ordinary differential equation was solved
numerically by the explicit embedded Runge-Kutta Prince-Dormand method
(8,9) \cite{Runge}.
The results were then compared with the usual method based on the five point stencil discretization of the differential operator with an application of the standard Octave method for finding the lowest eigenvalue of the resulting matrix \cite{octava}. Both approaches coincide, although the matrix approach seems to be less accurate in the vicinity of the mass threshold. 
\begin{figure}
\center
\includegraphics[width=9.5cm]{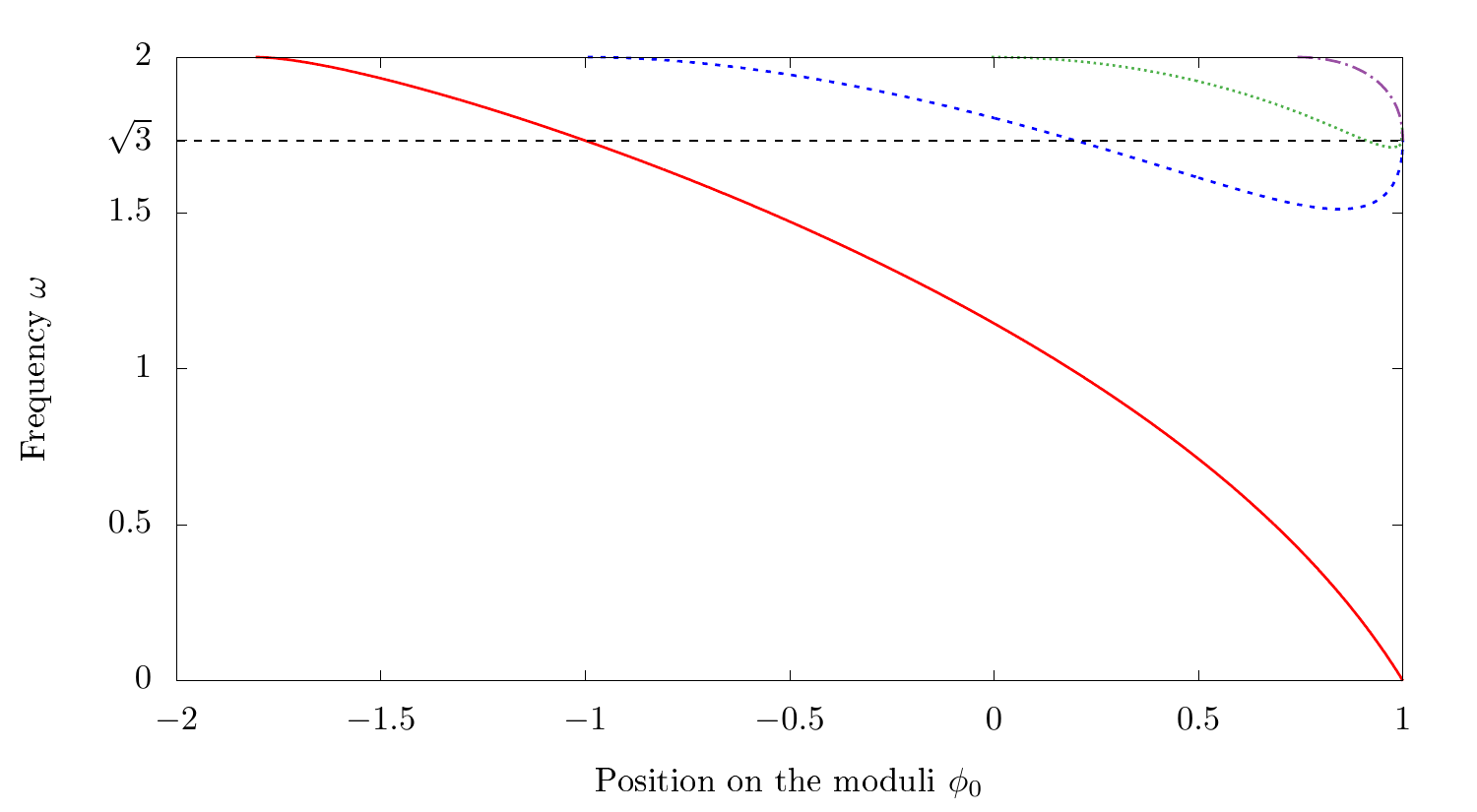}
\caption{
The spectral structure flow (bound modes) on the moduli space for the self-dual mode (\ref{L-sd}). There are altogether five frequencies, including the black line at $\omega =0$, corresponding to the zero mode (moduli). The dashed horizontal line denotes the frequency of the asymptotic shape mode of the free solitons.}
\label{modes}
\end{figure}

In Fig. \ref{modes} we display the mode structure as a function of the moduli space coordinate $\phi_0$. There are maximally four massive bound modes and one zero mode, which obviously generates the flow on the one-dimensional moduli space. The mass threshold is $\phi_0$ independent and is $\omega=2$. The observed normal modes may be best understood in the limit of widely separated solitons, i.e., in the limit $\phi_0 \to 1$. In this limit, both the kink and the antikink behave as independent solitons of the pure $\phi^4$ theory, therefore, each of them carries one zero mode and one shape mode with frequency $\omega = \sqrt{3}$. For an SAS pair at a finite separation ($\phi_0 <1$), the symmetric superposition of the two zero modes remains a zero mode (the moduli), whereas the anti-symmetric superposition transforms into a non-zero mode (the solid (red) line in Fig. \ref{modes}). Further, the symmetric and antisymmetric superpositions of the shape modes  turn into two non-zero modes emerging from $\omega = \sqrt{3}$ at $\phi_0=1$. The third mode emerging from there is a mode hosted by the impurity, and it starts precisely at $\sqrt{3}$ only for the particular choice (\ref{imp}) of the impurity. For other choices of the background field $\sigma$, both the number of modes hosted by the impurity and their 
frequencies change \cite{no-force-scatt}. We remark that the impurity mode (or wave function) is concentrated at the impurity, whereas the mode functions of the kink and antikink are concentrated on the kink and antikink, respectively. For a sufficiently separated SAS pair there is, therefore, no overlap between these mode functions and no interference of the impurity mode with excitations of the kink/antikink modes.

Probably the most important feature of the spectral structure flow is the disappearance of the bound modes into the continuous spectrum. This profoundly modifies the space of small perturbations and, as a consequence, affects the transition of energy stored  between kinetic and internal degrees of freedom. We underline that it is a very special property of the self-dual background field deformation that the mode structure is well defined at any intersoliton distance. Thus, we have the rare opportunity to analyze the relation between the excitation of bound modes and the soliton dynamics in a mathematically well defined way.

\subsection{Exciting the bound modes}
To understand the interplay of the kinetic (zero mode) and internal (bound modes) degrees of freedom, we consider a largerly separated kink-antikink pair with a bound mode exited as the initial state. Concretely, we excite the antisymmetric superposition of the shape modes. For sufficient initial separation, we can choose the known mode functions of the asymptotically free kink and antikink of the usual $\phi^4$ model for the initial state $\phi_{ in}$, 
 \begin{eqnarray}
 \phi_{ in}(x,t) &=& - \tanh\left(\gamma (x-x_0+vt) \right)  + \tanh\left(\gamma (x+x_0-vt) \right) -1\nonumber \\
 \nonumber \\
& & \hspace*{0.0cm}  - A \frac{\sinh \left(\gamma (x-x_0+vt) \right) }{\cosh^2 \left(\gamma (x-x_0+vt) \right) } \cos \omega \gamma (t + vx)  \nonumber \\
\nonumber \\
& & \hspace*{0.0cm}  - A \frac{\sinh \left(\gamma (x+x_0-vt) \right) }{\cosh^2 \left(\gamma(x+x_0-vt) \right) } \cos \omega \gamma (t-vx)  \label{init-pert}
\\ \nonumber 
\end{eqnarray} 
where the frequency $\omega^2= 3$. Further, $A$ is the amplitude of the excited mode (for the excitation of other modes we refer to \cite{no-force-scatt}). Such a superposition combines into a bound mode whose flow is given by the dotted (green) line in Fig. \ref{modes}. As the solitons approach each other during the collision, the frequency of the mode changes. It slowly rises and, importantly, it crosses the mass threshold at $\phi_0=\phi_{sw}\equiv-0.013$. 

\subsubsection{The spectral wall}
Exactly at the point $\phi_{sw}$ where the bound mode enters the continuous spectrum, a new phenomenon, called {\it spectral wall} (SW), occurs  \cite{spectral-wall, no-force-scatt}.  

A spectral wall is a kind of obstacle in the soliton evolution, where the SAS pair may form a stationary oscillating state. It emerges as a consequence of the transition of the bound mode into the continuous spectrum and shows up exactly at the point in moduli space $\phi_{sw}$ at which the corresponding mode crosses the mass threshold. As the moduli space coordinate can, to some extent, be translated into the SAS distance, the SW is effectively observed as a barrier in real (physical) space. Depending on the amplitude of the particular exited mode which is responsible for the SW, we have three possible scenarios. For a critical value of the amplitude, $A=A_{crit}$, the solitons are temporarily trapped at the SW distance, forming a (in principle infinitely) long-living stationary solution. For a less excited mode, $A<A_{crit}$, the solitons go through the wall and finally interact as is dictated by the moduli flow. In the case when the mode is excited more strongly, $A>A_{crit}$, the solitons are reflected back, where the reflection occurs sooner (at a larger intersoliton distance) as $A$ increases. Hence, a spectral wall acts as a filter which does not allow kinks to pass through if the pertinent mode is excited too much. Therefore, spectral walls are one of the leading factors governing the dynamics beyond geodesic flow in an SAS process in the limit where the static intersoliton force vanishes (self-dual process). 

\begin{figure}
\includegraphics[width=10.0cm]{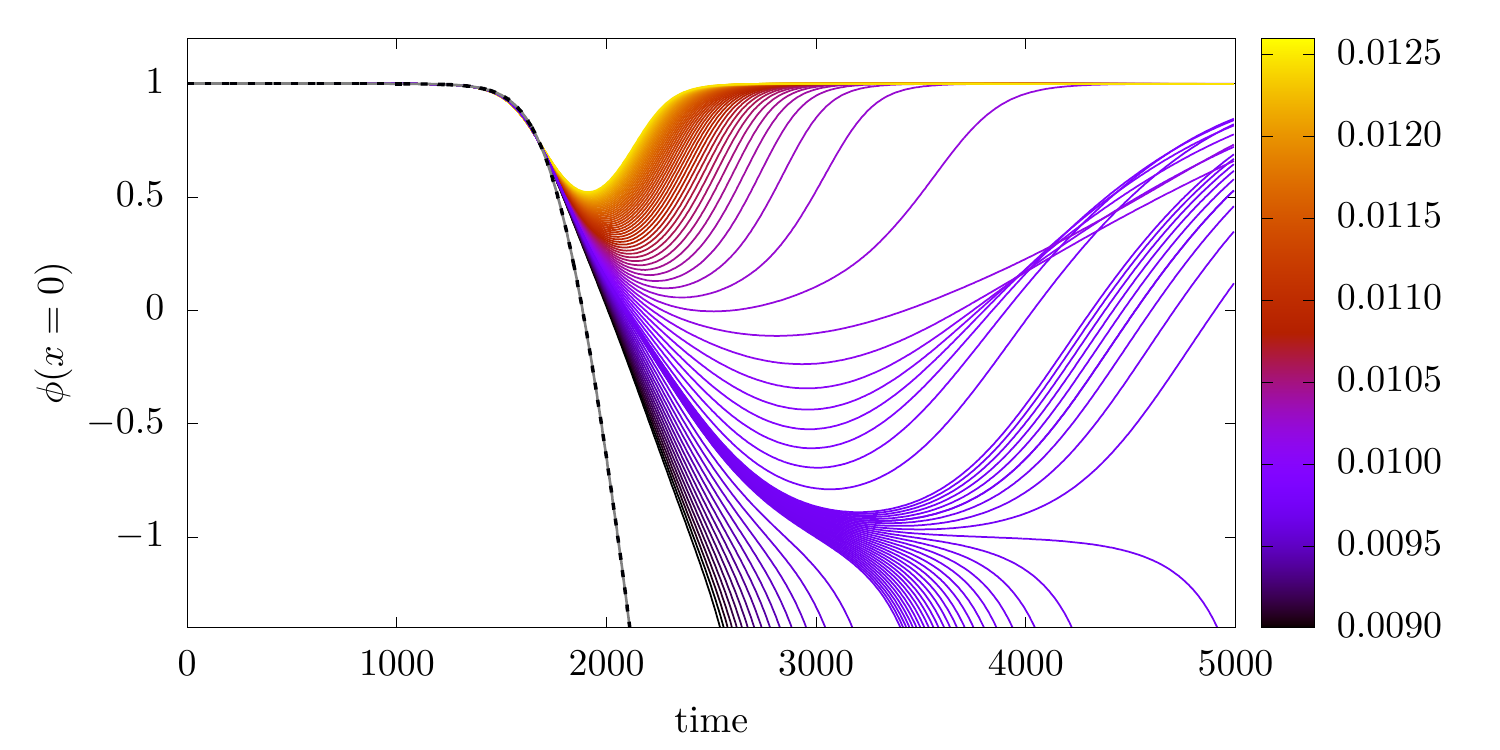}
\caption{Dynamics of $\phi (x=0,t)$ for different amplitudes of the mode (grey (color) scale) in the zero-static force limit. The SW at $\phi_0=\phi_{sw}=-0.013$ is hidden in the shadow of the vacuum wall (VW) located at $\phi_0=-1$. For comparison, we also show the case $A=0$ (no bound mode excited). The black dashed line shows the full numerical solution of Eq. (\ref{BPS-EL-eq}) with initial condition (\ref{init-unpert}), whereas the gray line shows the pure moduli space dynamics with the same initial condition. Both agree with high accuracy.   }
\label{wall-zero}
\end{figure}

In Fig. \ref{wall-zero} we show the evolution of $\phi(x=0,t)$ for the initial configuration (\ref{init-pert}) in the full Euler-Lagrange equation. Here $v=0.005$, $x_0=10$, and different colors represents different amplitudes $A$. It is clearly visible that the SAS trajectory can be reflected or transmitted through the SW. On the other hand, for this specific case the expected stationary solution at $\phi_0=-0.013$ (the position of this particular SW) does not form. The reason is that this SW is strongly affected by the vicinity of the vacuum wall (VW) at $\phi_0 =-1$ (the VW and its effects are explained below). The VW provides an additional repulsive force such that the expected stationary solution turns into a bounce. Our reason for choosing this particular SW is that even for a small breaking of self-duality the VW quickly disappears and the stationary solution related to the VW becomes clearly visible, see section 4. This allows us to demonstrate two phenomena---the fact that some SW can be partially hidden by the VW, and the fate of both the SW and the VW when self-duality is broken---with one example. For other SW which lead to clearly established stationary solutions already in the self-dual case we refer to \cite{no-force-scatt}.

Note that each bound mode is responsible for a different SW. In other words, a SW is a very {\it selective phenomenon} - a SW related to one mode is transparent for an SAS pair with another mode excited. 

Another important property of a spectral wall is that its position does not depend on the details of the initial configuration, e.g., the velocity of the incoming solitons. It is always located at the same spatial point. Therefore, it looks like a stiff {\it infinitely thin} wall. 

Unfortunately, although the position of a SW (the distance between kink and antikink) can be easily predicted, we were not able to compute the value of the critical amplitude analytically. The main problem is that any effective model, which includes the normal modes in addition to the moduli space (the zero mode), collapses exactly at the point where a SW shows up, i.e., where a mode crosses the mass threshold, because the corresponding mode function is no longer normalizable there \cite{spectral-wall}. 

\vspace*{0.2cm} 

In addition, higher order spectral walls have been also found. They are triggered by a similar mechanism. In this case, it is a transition of a higher harmonics (of a normal mode) into the continuum spectrum which is responsible for its appearance. The main effect caused by a higher spectral wall is a huge radiation burst \cite{no-force-scatt}. 
\subsubsection{The vacuum wall}
There is another phenomenon in the zero-static force limit version of the $\phi^4$ model which profoundly affect the SAS dynamics, namely, the {\it vacuum wall} (VW). It is again an obstacle in the soliton dynamics located at $\phi_0=-1$, corrresponding to the vacuum solution $\phi(x; \phi_0=-1)=-1$, which explains the origin of the name. Whenever the field passes the vacuum it may be affected by the vacuum wall and, for a critical value of an excitation, another stationary solution can be formed. Again, for a less excited SAS pair the solitons pass the VW and behave as predicted by the geodesic flow, while for more excited pairs the solitons are reflected backwards, see Fig. \ref{wall-zero}. However,  contrary to a SW the VW is not a selective phenomenon, related to a specific internal mode. Any sufficiently large excitation of the SAS pair will lead to the formation of this stationary solution. The VW is explained within the higher order perturbation theory \cite{no-force-scatt}, but a heuristic argument is like follows. An initial perturbation is added at both, infinitely separated solitons, for $\phi_0 \to 1$,  when they behave as usual solitons in the $\phi^4$ model. During the evolution we reach the $\phi=-1$ vacuum solution with the initial perturbations on top of it. But the $\phi_0=-1$ solution is a very sensitive point in configuration space, because the corresponding vacuum configuration has zero energy density. As a consequence, any local perturbation around it may grow with almost no cost in energy and create an SAS pair from the vacuum ($\phi_0$ grows) or lead to the formation of the infinitely deep bump ($\phi_0$ decreases to minus infinity). In this situation, higher order nonlinearities decide which direction is realized during the actual evolution. 

The vacuum wall is a stronger effect than the spectral wall in the sense that if a SW is located too close to the VW (in the moduli space), then the formation of the stationary solution due to the SW may be impossible. This effect, called the shadow of the VW, is clearly seen in Fig. \ref{wall-zero}. The $\phi_0$ trajectory is slightly flattened at the position of the spectral wall but then it is repelled due to the interaction with the vacuum wall, and a stationary solution never forms. One can say that the spectral wall is hidden in the shadow of the vacuum wall.

\section{Introducing the static intersoliton force}
\subsection{The prescription}
The unique, special feature of the self-dual background field deformation of a solitonic process in a given field theoretical model is that there is no static force between the solitons. To relate the process in question to a realistic (non-selfdual or non-BPS) counterpart, one has to switch on such forces. Basically, from a qualitative point of view, any breaking of the BPS property is equivalent, i.e., it introduces an intersoliton static force. As a consequence, the resulting theory should qualitatively reproduce the main features of the considered SAS process in the original, $L[\phi]$ theory, regardless of the particular way of breaking the self-duality. 

Among the infinitely many possible ways to break the BPS property of an SAS process in the $L[\phi, \sigma]$ model, there is one particularly simple one. Namely,  
\begin{equation} \label{L-nsd}
L[\phi, \sigma; \epsilon]=(1-\epsilon^2) L[\phi,\sigma]+ \epsilon^2 L [\phi]
\end{equation}
where $\epsilon \in [0,1]$ is a BPS breaking parameter, allowing to interpolate between the zero static force limit i.e., the self-dual background field model $L[\phi, \sigma]$ ($\epsilon=0$) and the original theory 
\begin{equation}
 L[\phi]=\frac{1}{2} \int_{-\infty}^\infty  dx \left[ \phi_t^2- 
\phi_x^2  - W'^2    \right] 
 \end{equation}
 ($\epsilon=1$).
After integrating by parts and dropping the complete
derivative part ($\frac{d}{dx}(\sigma W\phi)$), the self-dual Lagrangian can be written as as standard Lagrangian with a background-dependent potential, 
\begin{equation}
L[\phi, \sigma]=\int_{-\infty}^\infty  dx \left(  \frac{1}{2}\phi_t^2  - \frac{1}{2}
\phi_x^2 -\frac12\sigma^2W'^2+\sigma_xW \right) .
\end{equation}
This results in the full Lagrangian 
\begin{equation}
 L[\phi, \sigma; \epsilon]= \int_{-\infty}^\infty  dx \left( \frac{1}{2}\phi_t^2  -  \frac{1}{2} 
\phi_x^2 - U \right)
\end{equation}
with the field-theoretic potential
\begin{equation} \label{U-pot}
U =
-(1-\epsilon^2)\sigma_xW+\frac12 W'^2\left[(1-\epsilon^2)\sigma^2+\epsilon^2\right] .
\end{equation}
The equations of motions are
\begin{equation} \label{field-eq}
 \phi_{tt}-\phi_{xx}-U'(\phi) = 0,
\end{equation}
where, using the explicit expression (\ref{superpot}) for $W$,
\begin{equation}
 U'=(1-\epsilon^2)(1-\phi^2)\sigma_x - 
2\phi(1-\phi^2)[(1-\epsilon^2)\sigma^2-\epsilon^2] .
\end{equation}
Note that this model has the same mass threshold, $\omega^2=4$, for any $\epsilon$. Furthermore, since for $x \rightarrow \pm \infty$ the impurity tends to $\pm 1$, we recover the usual $\phi^4$ model in these limits. Hence, asymptotic states are exactly the $\phi^4$ kink and antikink. Also the kinetic term is independent of $\epsilon$. 

In the subsequent part of the paper we will analyze this theory in the regime where $\epsilon \ll 1$, that is, where the intersoliton force is weak. 
  
\subsection{The unstable manifold}
For $\epsilon \ll 1$ we can still treat the BPS configuration (\ref{sol-sd}) with a time-dependent $\phi_0$ as a good approximation to the true solution for small velocities. 
If we insert it into the full non-selfdual lagrangian (\ref{L-nsd}) then, in addition to the kinetic term which induces the moduli space metric, we get a static term which may be interpreted as an effective potential $V(\phi_0)$, i.e., 
\begin{equation} \label{L-eff}
L_{\rm eff}=\frac{1}{2}M (\phi_0) \dot{\phi}_0^2 - \epsilon^2 V(\phi_0),
\end{equation}
where
\begin{equation}
V(\phi_0) \equiv \int dx \frac{1}{2} \left( [\phi ' (x;\phi_0)]^2 + [1-\phi^2 (x;\phi_0)]^2 \right) .
\end{equation}
Formally we may perform this substitution for any value of $\epsilon$, but the resulting effective lagrangian can provide a reasonable approximation only for sufficiently small $\epsilon$, for which the agreement is excellent, see Figs. \ref{wall-eps} and \ref{wall-eps-2} below. This moduli space with a nontrivial potential is usually referred to as an unstable manifold \cite{Manton-1}. The self-dual breaking part modifies the geodesic flow by the appearance of a drag force due to the effective potential $V(\phi_0)$.

The $x$ integral defining $V(\phi_0)$ can be done exactly, resulting in
(again, we display the expression for $V(a)$ instead of $V(\phi_0)$ to obtain a simpler expression)
\begin{equation}
V(a) =\frac{1}{3(1+a)^3}\left( 3 + 4 a (1 + 2 a(3 + a) ) +  3\frac{-1 - 2 a + 4 a^2}{\sqrt{a(1+a)}} \mbox{arctanh} \sqrt{\frac{a}{1+a}} \; \right) . \label{eff_pot}
\end{equation}
We plot this effective potential in Fig. \ref{effectiv}, right panel, curve $V(\phi_0)$.

The BPS breaking term strongly modifies the geodesic dynamics for $\phi_0<-1$,  if compared with the BPS (zero-force) limit, because the effective potential grows rapidly there, inducing a strong force towards larger values of $\phi_0$. In other words, a repulsive core emerges for $\phi_0 <-1$. This provides a mechanism which prevents the solutions of the effective lagrangian (\ref{L-eff}) from developing a too large negative bump. Indeed, any trajectory on the unstable manifold is now reflected back at $\phi_0^*$, which we call the reflection point. Its position can be obtained from the first integral
\begin{equation}
\frac{1}{2} M(\phi_0) \dot{\phi}_0^2 + \epsilon^2 V(\phi_0) = E
\end{equation}
where $E$ is the initial energy. Indeed,  $\phi_0^*$ obeys the relation 
\begin{equation}
\frac{1}{2} M(\phi_0^{in}) (\dot{\phi}_0^{in})^2 + \epsilon^2 V(\phi_0^{in}) = \epsilon^2 V(\phi_0^*) 
\end{equation}
where $\phi_0^{in}$ and $\dot{\phi}_0^{in}$ is the initial position on the moduli and its velocity. Assuming that the initial state consists of the infinitely separated kink and antikink with velocity $v$ we find that $\phi^{in}_0 \rightarrow 1$  and $\dot{\phi}_0^{in}=  -2v\gamma (1-\phi_0^{in})$ (here $\gamma=1/\sqrt{1-v^2}$). Thus, the reflection point can be read off from the algebraic equation (see Appendix B)
\begin{equation}
V(\phi_0^*)=\frac{4}{3}\left(\frac{ v^2\gamma^2}{\epsilon^2} +2\right). \label{core}
\end{equation}
As expected, the position of the reflection point depends on the initial velocity $v$ and the self-duality breaking parameter $\epsilon$. As $\epsilon$ tends to 0 or $v$ increases, we can climb higher on $V$, that is, go further in the moduli space towards more negative $\phi_0$. 

Note that the effective potential has a local maximum at $\phi_0^b=0.89167$ (see the small insert in Fig. \ref{effectiv}, right panel). This means that kinks in the initially infinitely separated SAS pair repel each other. After crossing this little barrier, they start to attract. The barrier is very small, because $V(\phi_0^b)-V(\phi_0=1)=(2.67029-8/3)=0.00363$. Therefore, at least for a weak self-duality breaking (small $\epsilon^2$), it is very easy for boosted solitons to climb over the potential barrier. 
 
 For the initial configuration (\ref{init-unpert}), where no massive modes are excited, the unstable moduli description reproduces the full dynamics very well. 

\section{The fate of the spectral and vacuum walls }
\begin{figure}
\includegraphics[width=8.0cm]{eps0.pdf}
\includegraphics[width=8.0cm]{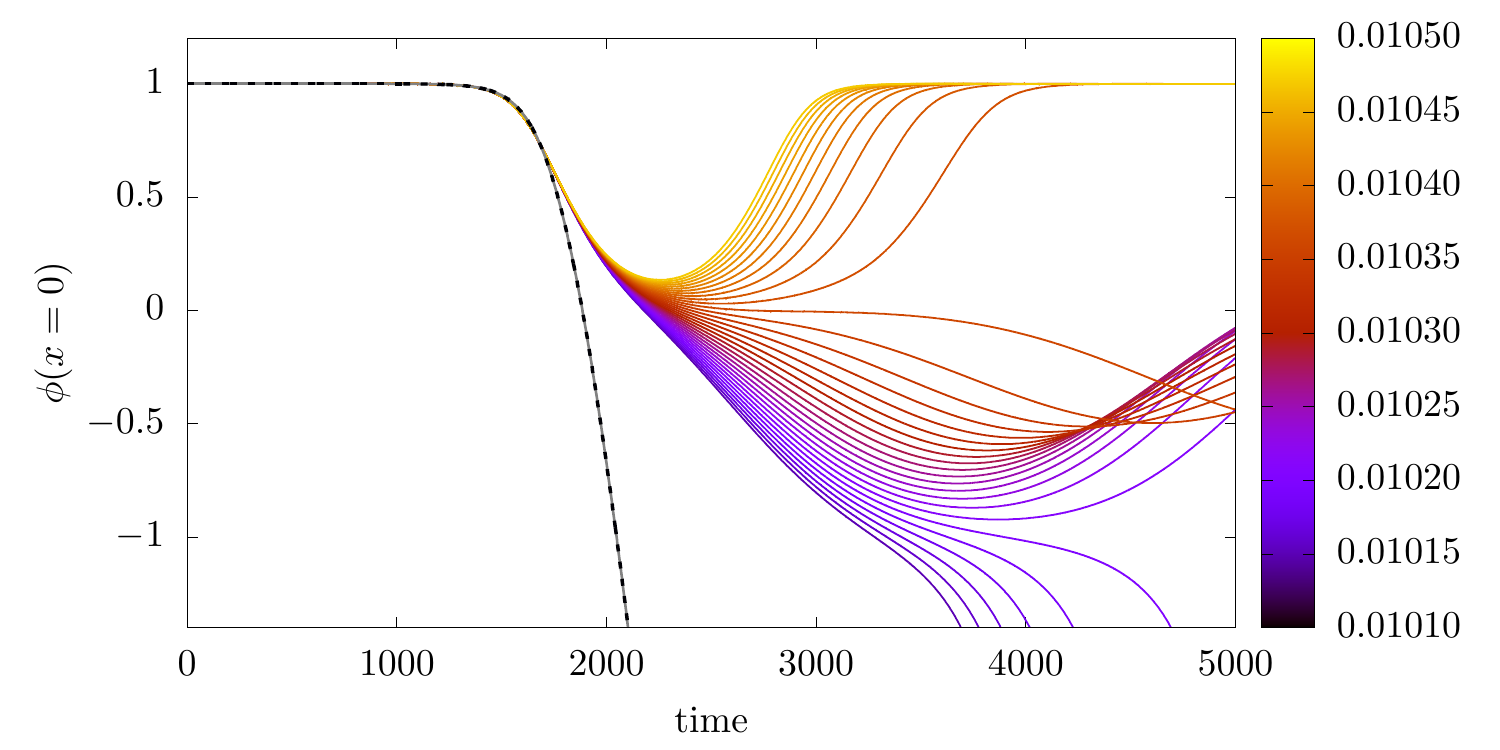}
\caption{Dynamics of $\phi_0$ for different amplitudes of the mode (grey (color) scale) . Left: $\epsilon=0$ (the BPS case i.e., the zero force limit) - the SW at $\phi_0=-0.013$ is hidden in the shadow of the VW. Right: $\epsilon=0.001$ - the barrier is very well visible and the VW is still present. For comparison, we also show the case $A=0$ (no bound mode excited). The black dashed line shows the full numerical solution of Eq. (\ref{field-eq}) with initial condition (\ref{init-unpert}), whereas the grey line shows the  effective dynamics on the unstable manifold resulting from (\ref{L-eff}), with the same initial condition. Both agree with high accuracy. }
\label{wall-eps}
\includegraphics[width=8.0cm]{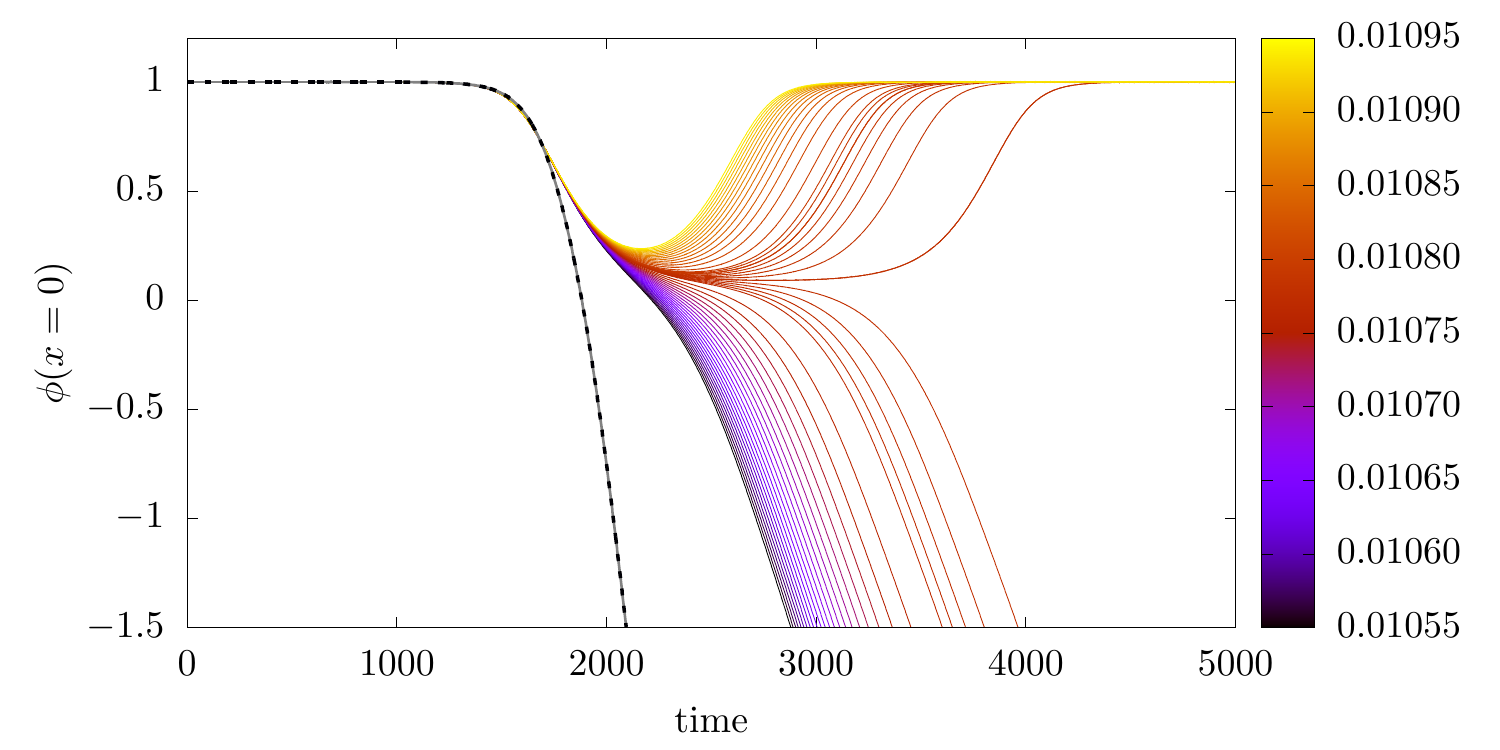}
\includegraphics[width=8.0cm]{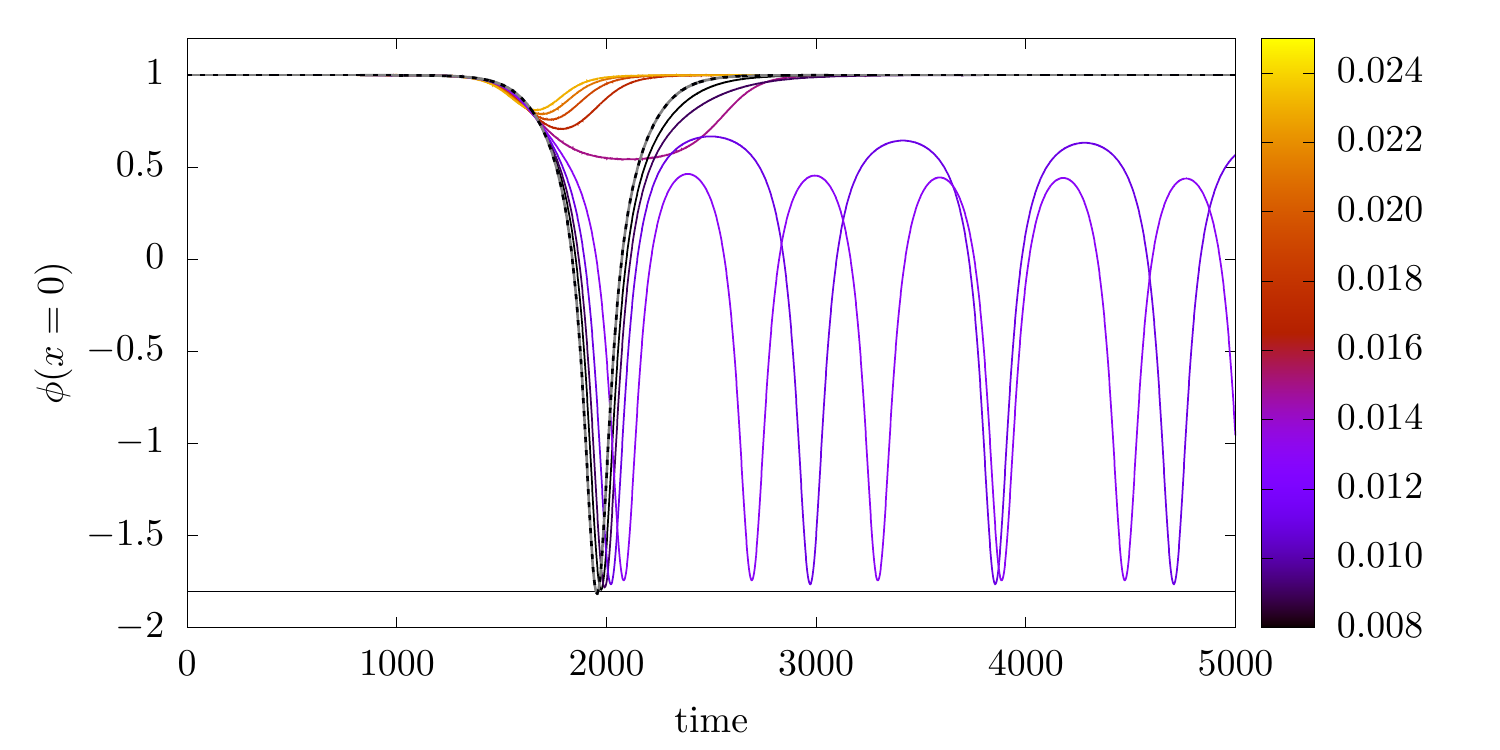}
\caption{Dynamics of $\phi_0$ for different amplitudes of the mode (grey (color) scale). Left: $\epsilon=0.002$ - the barrier moves up while the VW disappears. Right: $\epsilon=0.01$ - the barrier moves further up. The repulsive core is visible and $\phi^*_0=-1.804$ is the corresponding reflection point (horizontal line). Note the bouncing solutions trapped between the core and the barrier. For comparison, we also show the case $A=0$ (no bound mode excited). The black dashed line shows the full numerical solution of Eq. (\ref{field-eq}) with initial condition (\ref{init-unpert}), whereas the grey line shows the effective dynamics on the unstable manifold resulting from (\ref{L-eff}), with the same initial condition. Both agree with high accuracy. }
\label{wall-eps-2}
\end{figure}

To study the fate of the spectral walls and the vacuum wall in the $\epsilon \ll 1 $ regime we, again, consider an excited largely separated kink-antikink pair (\ref{init-pert}). In Figs. \ref{wall-eps} and \ref{wall-eps-2} we display the dynamical evolution of $\phi_0$ for such an initial state. That is to say, the initial configuration (\ref{init-pert}) is inserted into the full, time-dependent Euler-Lagrange equations resulting from the lagrangian (\ref{L-nsd}) which are then evolved numerically by converting them into a system of ordinary differential
equations using the 5 point stencil spatial discretization and the fourth order
symplectic method for time integration.
The initial velocity and position of the initial configuration are always $v=0.005$, $x_0 = 10$, whereas the different values for $\epsilon$ are provided in the figure captions. Finally, the different values for the amplitude $A$ of the normal mode are indicated by the grey (color)  scheme.
The reason why we plot the time evolution of $\phi (x=0,t)\equiv \phi_0 (t)$ 
in Figs. \ref{wall-eps} and \ref{wall-eps-2} is that this variable---which we call the "trajectory" of the solution---provides a useful representation of the corresponding solution $\phi (x,t)$. Concretely, $\phi_0 (t) \sim 1$ corresponds to a well-separated SAS pair, $\phi_0 (t) \sim 0$ to a configuration where the individual solitons have dissolved, and $\phi_0 (t) \sim -1$ to a transition through the vacuum $\phi = -1$. Finally, $\phi_0 (t) < -1$ corresponds to the formation of the negative bump.

Now we switch on the BPS breaking term by taking $\epsilon > 0$, i.e., introducing a non-zero static intersoliton force. We begin with a very small velocity $v=0.005$. The first effect is that the vacuum wall is less important as $\epsilon$ increases, and very quickly this wall disappears completely. The reason for that is the following. The vacuum wall in the BPS case exists as a consequence of the zero mode. That is to say, when a BPS solution passes through the vacuum $\phi =-1$, an arbitrarily small perturbation is sufficient to form an SAS pair. After the breaking of the BPS property (self-duality), the vacuum $\phi=-1$ is still a solution. However, fluctuations now cost energy and, therefore, small perturbations do not grow arbitrarily (forming SAS pairs), but either are localized as oscillations around the vacuum or travel deep into the negative values of $\phi_0$ where they meet the repulsive core. In other words, the force induced by the non-BPS part takes the trajectory through $\phi_0=-1$, thus eradicating the VW effect. This occurs, in fact, very quickly. The VW exists for $\epsilon=0.001$ (Fig. \ref{wall-eps} right panel) but already for $\epsilon=0.002$ it completely disappears (Fig. \ref{wall-eps-2} left panel). 

Secondly, the reduced importance of the vacuum wall and, as a consequence, its shadow, makes the stationary solution very well visible, see for example Fig. \ref{wall-eps} (right panel) where the dynamics for $\epsilon=0.001$ is presented. The vacuum wall is still there, but the stationary solution related to the SW is now clearly visible. 

\begin{figure}
\includegraphics[width=7.5cm]{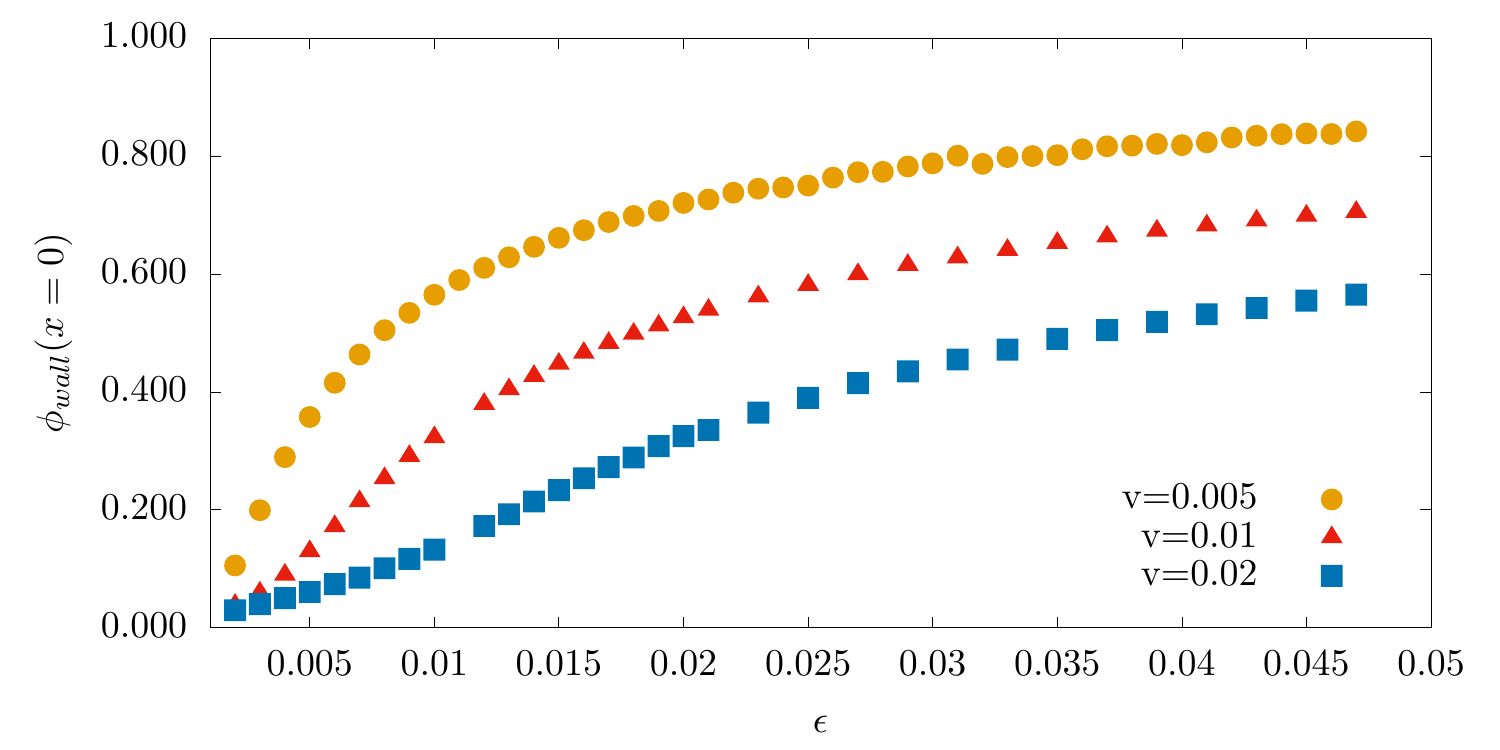}
\includegraphics[width=7.5cm]{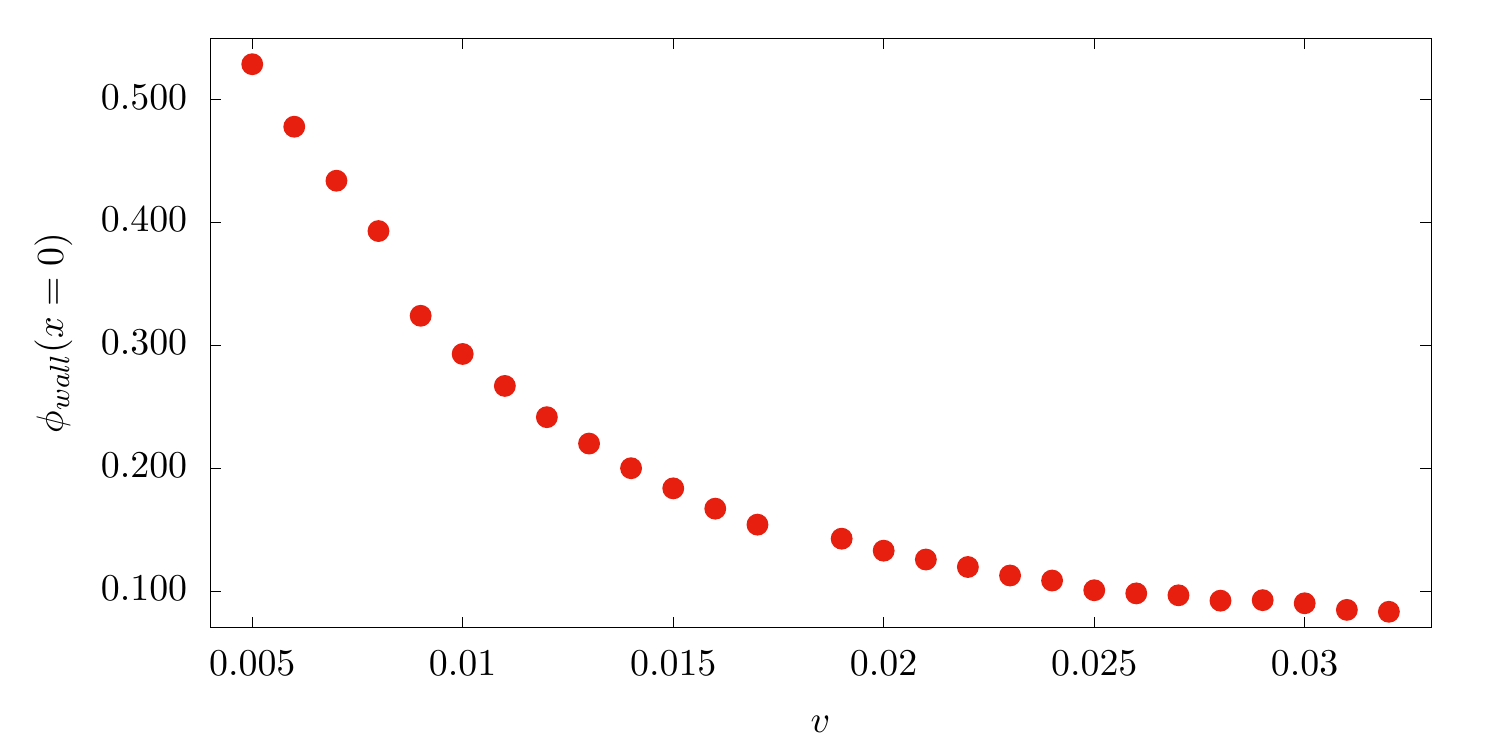}
\caption{Position of the barrier (stationary solution).  Left: as a function of $\epsilon$. Right: as a function of velocity ($\epsilon=0.01$).}
\label{wall}
\end{figure}

Importantly, the position of the stationary solution in moduli space, which we call a barrier, rapidly moves towards bigger values of $\phi_0$ (see Fig. \ref{wall-eps-2} right panel).  Observe that, due to the weak BPS (self-duality) breaking, $\phi_0$ may still be used as a relevant coordinate on the unstable manifold, giving a good insight into the dynamics of the system. In fact, even for very small values of the parameter $\epsilon$, the barrier varies significantly, e.g.,  from $\phi_0=-0.013$ for $\epsilon=0$ to $\phi_0=0.424$ for $\epsilon=0.008$, see Fig. \ref{wall} (left panel, dots). Now the SW forms for a sufficiently separated SAS pair, where the individual kink and antikink can still be identified. 

For higher values of $\epsilon$, the repulsive core is also well visible, Fig. \ref{wall-eps-2} (right panel), and the position of the reflection point is captured by our effective model analysis with a very good accuracy, see eq. (\ref{core}). Note that, for this case, also bouncing solutions are present. They describe an SAS pair trapped between the core and the barrier where the stationary solution is formed. 

\section{Thick spectral wall}

A closer look at the barrier in the non-BPS regime reveals even more fascinating findings. 
In contrast to the BPS (zero-force) limit, the position of the stationary solution (the barrier) now significantly depends on the velocity of the incoming solitons. Indeed, it goes toward smaller $\phi_0$ as $v$ increases, see Fig. \ref{wall} right panel. This means that the barrier, which coincided with the fixed position $\phi_{sw}$ of the (thin) spectral wall in the zero static force case (at $\epsilon=0$), now (for $\epsilon \neq 0$) occurs at some position $\phi_0 > \phi_{sw}$ inside a finite region, which we call a {\it thick spectral wall}.
Further, this thick spectral wall shows a certain "stiffness", i.e., a certain resistance against a too close approach of the kink-antikink pair.
Specifically, a bigger stiffness means that the position of the stationary solution
(the barrier) for a fixed initial velocity is located at a larger
$\phi (x=0,t)$, i.e., when the solitons are further away from each other.
When the initial velocity grows, the resulting solution compresses the wall more strongly, approaching smaller values of $\phi_0$. On the other hand, the bottom of the thick spectral wall, which is located at the position $\phi_{sw}$ of the original thin SW, apparently does not move. Hence, at some point even a large change of $v$ changes the position of the stationary solution only very weakly. 

\begin{figure}
\center \includegraphics[width=12.0cm]{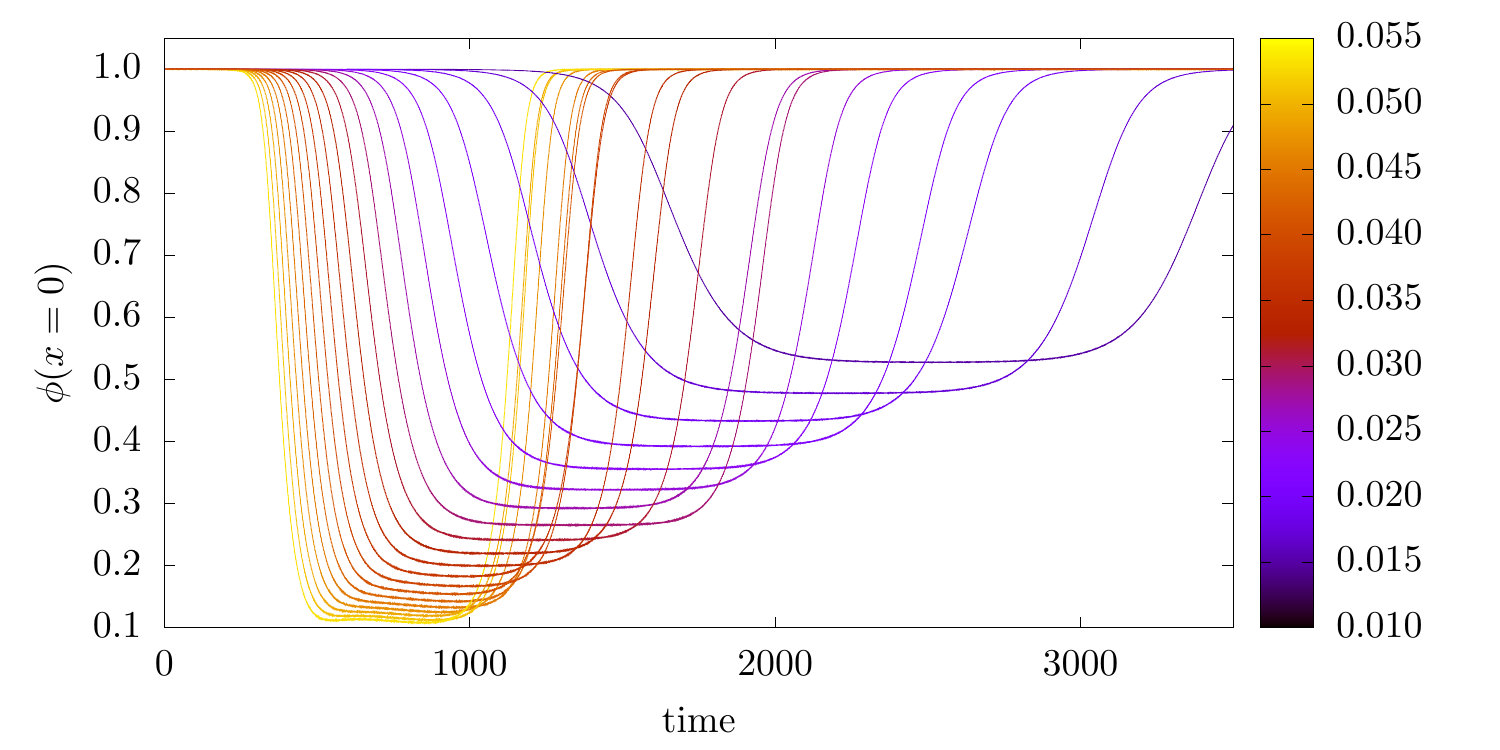}
\caption{Thick spectral wall for $\epsilon=0.01$. Dynamics of $\phi_0$ for different initial velocities of the incoming solitons (different shades of gray (colors)) with the stationary solution (the flat region of the trajectory) frozen at the barrier.}
\label{sponge}
\end{figure}

An example is shown in Fig. \ref{sponge} where, again, we evolve the full system with initial condition (\ref{init-pert}), now for $\epsilon = 0.01$, $x_0 = 10$, and for different initial velocities. 
For each initial velocity, we only show the particular solution (the particular value of the amplitude $A$) which freezes on the thick wall, i.e., which remains stationary for a sufficiently long time. 
For larger velocities, $\phi_0 (t)$ gets closer to the bottom of the thick spectral wall. It turns out, however, that the numerical calculation of the SAS scattering is difficult for very large velocities, therefore we study the approach to the bottom of the thick wall (the original thin SW) by a slightly different method, see Fig. \ref{saddle} below.  We further find that, if we vary $\epsilon$ instead of $v$,  the stiffness of the thick spectral wall increases with the BPS breaking parameter $\epsilon$. 

\begin{figure}
 \includegraphics[width=15.0cm]{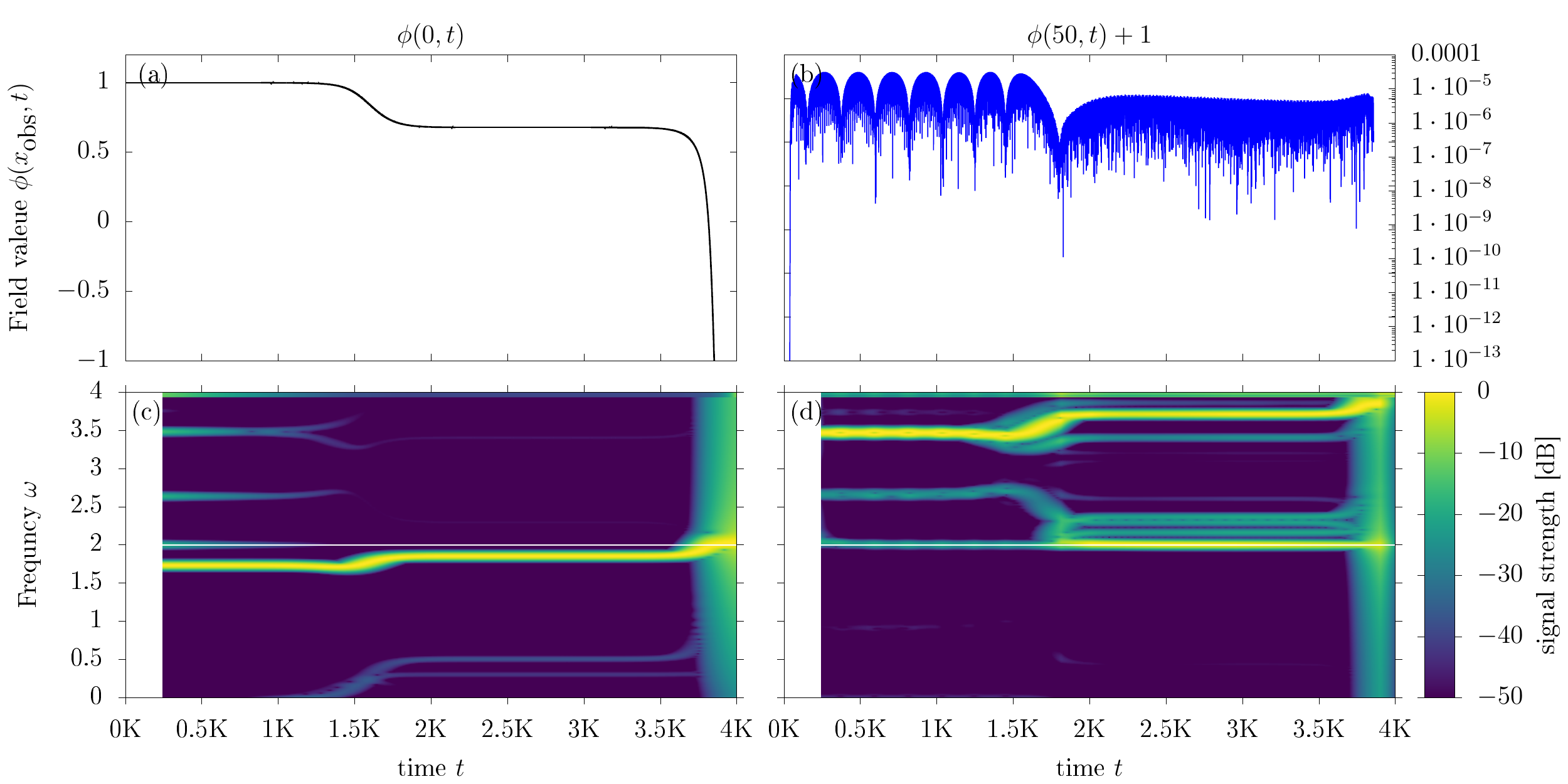}
\caption{Solution frozen on the barrier in the thick spectral wall for $\epsilon=0.02$. Upper left: dynamics of $\phi_0$. Lower left: strength of the  excited modes at $x=0$. Upper right: radiation at $x=50$. Lower right: excited modes at $x=50$.  Here $x_{\mbox{{\tiny obs}}}$ represents the two positions $x=0$ or $x=50$ where the field is observed.}
\label{spectral-0.02}
\end{figure}

As we know, in the zero static force limit the thin spectral wall occurs when a mode enters the continuum. However, when we depart from this extreme BPS regime, we find that the barrier happens {\it before} the excited mode enters the continuum, e.g., for $\phi_0 >\phi_{sw}$, see Fig. \ref{spectral-0.02}. 
In more detail, in the upper left panel of Fig. \ref{spectral-0.02}, we show the time evolution of $\phi_0 (t)$ for  $\epsilon = 0.02$, for the initial configuration (\ref{init-pert}) with parameter values $v = 0.0050$, $A= 0.018760$ and initial soliton positions $x_0 = 10.0$. 
Once the two solitons get sufficiently close such that $\phi_0 (t)$ deviates notably from $\phi_0 =1$,
the transition to the stationary solution occurs. The solution remains at this stationary position from $t\sim 1700$ until $t\sim 3600$, and then further decays towards the vacuum and the negative bump. In the upper right panel, we show the field far away from the interaction region, $\phi (x=50,t)$ (more precisely, its deviation from the asymptotic vacuum value $\phi =-1$), which essentially reveals the radiation carried away in the process. Finally, in the lower panels, we show the corresponding temporal Fourier transforms (more precisely, the Gabor transforms \cite{gabor}), which disclose information about the participating excited modes and their strengths. 
The fact that the stationary solution happens before the (thin) spectral wall implies that the mechanism behind the thick spectral wall phenomenon must be  a {\it modification} of the thin SW mechanism. 

\subsection{Perturbative analysis}
It is quite difficult to solve the equation of motion for a full dynamical
process such as a SAS collision. But by using perturbative techniques, we can
show that there exist some special, stationary solutions. 
In the case of the BPS model, the thin SW itself provides an example of a stationary solution.
In this section, we show that oscillating perturbations (excited modes) can
move the position of the barrier. This gives rise to the appearance of the
thick SW phenomenon.

The main idea is that nonlinear contributions coming from the excitation of
the first excited bound mode $\phi_1$ \textit{cancel} the acceleration resulting from the
effective potential (\ref{eff_pot}). This balance leads to the formation of the
unstable stationary solution.
Concretely, we search for an approximated solution $\Phi$ of the form 
\begin{equation}
\phi \approx \Phi \equiv \phi (x,\phi_0) + A_0\eta_0 +A_1\phi^{(1)} + A_1^2\phi^{(2)} , \label{pert}
\end{equation}
and we want to investigate under which conditions this approximated solution can be stationary.
Here $ \phi(x; \phi_0)$ is the self-dual (BPS) solution (\ref{sol-sd}). We also add the mode $\eta_0(x)$ which corresponds to the zero mode in the self-dual case and turns into an unstable mode with $\omega_0^2 <0$ for the non-self-dual model ($\epsilon >0$). This mode
produces the flow on the unstable manifold, induced by the force due to the effective potential  (\ref{eff_pot}). Further, we add the excited bound mode $\phi^{(1)} =\phi_1= \cos (\omega_1 t) \eta_1(x)$ (here the upper index characterizes the perturbative order, while the lower index denotes the first excited mode). 
In this section,
we assume that all the bound modes are properly normalized, so that
$\int_{-\infty}^{\infty} \eta_i^2=1$ (note that in formula (\ref{init-pert}) we used
unnormalized modes for initial conditions). 
Finally, $A_0$, $A_1$ are the amplitudes of the two modes, respectively, and $\phi^{(1)}$  and $\phi^{(2)}$ are the first and  second order
corrections in $A_1$.

That is to say, Eq. (\ref{pert}) implies an expansion up to second order in $A_1$ and up to first order in $A_0$. The reason for the perturbative expansion in $A_1$ is that this coefficient is small for small $\epsilon$ (see Eq. (\ref{A1}) below) and that we have to go to second order to see the effect we want to show. The reason for an expansion up to first order in $A_0$ is related to the fact that $A_0 \eta_0$ is the infinitesimal (or first order) generator of the flow on the unstable manifold (or on the moduli space for $\epsilon=0$). To find a strictly stationary solution requires $A_0 =0$, and this is also a sufficient condition, because the flow generated by a zero vector field is zero. It is, therefore, sufficient to expand only up to first order in $A_0$. 

It turns out that requiring $A_0 =0$ is too restrictive. Instead, we shall find solutions of the type $A_0 \sim \epsilon^2 \cos (\omega t)$ which describe tiny oscillations about the stationary solution with a certain frequency $\omega$. As $A_0$ now is no longer strictly zero, higher orders in $A_0$ should, in principle, be included (i.e., the full, integrated flow should be considered). It can be expected that, after a sufficiently long time, higher order effects in $A_0$ will accumulate and destroy the (almost) stationary behavior, and our numerical calculations show that this is indeed what happens. But the almost stationary trajectory can, nevertheless, be clearly seen for rather long time intervals.

For the concrete calculation, we have to insert the approximation (\ref{pert}) into the full Euler-Lagrange equation (\ref{field-eq}). 
Expanding order by order in $A_1$, we get  at $\mathcal{O}(A^0_1)$ 
\begin{equation}
\phi^{(0)}_{tt}-\phi^{(0)}_{xx} + U' (\phi^{(0)})=0 .
\label{zero-ord-A1}
\end{equation} 
In the BPS case, this already leads to a stationary solution---in fact, to a static one, because $\omega_0 =0$. 
In the non-BPS model, there exist only three static solutions describing the vacuum, a pair of infinitely separated kinks, or an unstable solution related to the local maximum of the effective potential $V(\phi_0)$ (see Fig. 1, right panel). Other solutions are not static and accelerate due to the effective potential. Assuming now $\phi^{(0)}=\phi (x,\phi_0)  + A_0(t)\eta_0(x)$ as explained, and expanding $U'$ up to linear order in $A_0$, equation (\ref{zero-ord-A1}) becomes
\begin{equation}
\ddot{A}_0\eta_0 + A_0 H \eta_0= \phi'' (x,\phi_0) -U'
\end{equation} 
where 
\begin{equation}
H\equiv -d^2/dx^2 + U'' ,
\end{equation} 
and $U'\equiv U'(\phi (x,\phi_0))$ and $U''\equiv U''(\phi (x,\phi_0))$ are evaluated at the BPS solution $\phi (x, \phi_0)$. Using the fact that,  for $\epsilon >0$, $\eta_0$ is the unstable eigen-mode of $H$ with $\omega_0^2 <0$ and projecting the equation on $\eta_0$, we find
\begin{equation}
\ddot{A}_0  + \omega^2_0 A_0 = \int dx \eta_0 (\phi''(x,\phi_0)- U').
\end{equation} 
For the BPS, zero-static force case, the r.h.s. vanishes and $\omega_0=0$ ($\eta_0$ is a zero mode). In the non-BPS case, the general solution is
\begin{equation}
A_0 = c_+e^{|\omega_0| t} + c_- e^{-|\omega_0| t} + c
\end{equation}
where $c_+$ and $c_-$ are integration constants and 
\begin{equation}
c = \omega_0^{-2} \int dx \eta_0 (\phi''(x,\phi_0)- U').
\end{equation}
So even for $c_+ = c_- =0$ it holds that $A_0 = c \not= 0$, and a stationary solution cannot exist.

The first order correction is simply given by the bound mode $\phi_1$, 
\begin{equation}
\phi^{(1)}=\phi_1 = \frac{1}{2} \eta_1(x) e^{i\omega_1 t} +c.c.
\end{equation} 
and, therefore, shows a purely oscillatory behavior. 
Here, the eigenmode profiles $\eta_i$ and frequencies $\omega_i$  are obtained from the eigenvalue equation $H\eta_i = \omega_i^2 \eta_i$. 

Finally, the second order leads to an inhomogeneous equation
\begin{equation}
\phi^{(2)}_{tt}-H \phi^{(2)} = -\frac{1}{2}U''' \cdot (\phi^{(1)})^2 .
\end{equation} 
Taking all this information into account, the full equation up to second order projected onto the unstable mode $\eta_0$ reads
\begin{eqnarray}
\ddot{A}_0  + \omega^2_0 A_0 &=& \int dx \eta_0 (\phi''(x,\phi_0) - U') \nonumber  \\&-& \frac{1}{4}A_1^2 (1+\cos 2\omega_1 t) \int dx \eta_0 U'''\eta^2_1 , \label{full-2nd-order}
\end{eqnarray} 
where, again, $U'''=U'''(\phi (x,\phi_0))$ is evaluated at the BPS solution.
Thus, the critical amplitude $A_{1,{\rm c}}$ of the oscillational mode $\phi_1$ which cancels the acceleration generated by the effective potential is
\begin{equation}
A_{1,{\rm c}} (\phi_0)=2\sqrt{\frac{ \int dx \eta_0 (\phi''(x,\phi_0) - U')}{ \int dx \eta_0 U'''\eta^2_1}} \label{A1}
\end{equation}
which is proportional to $\epsilon$ in leading order. 
This equation gives, for  fixed $\epsilon$, a relation between the amplitude of the excited mode which leads to the stationary solution, i.e., the appearance of the barrier in the thick SW, and its position $\phi_0$ along the unstable manifold.   
 There is also a non-vanishing non-homogenous part, corresponding to the $\cos 2\omega_1 t$ term at the right-hand-side of (\ref{full-2nd-order}). This term, however, does not destabilize the solution for a rather long time but generates small oscillations with frequency $2\omega_1$. 
 In fact, we have seen these small oscillations when we zoomed on the stationary
trajectory frozen on the thick SW.
 \begin{figure}
\center \includegraphics[width=12.0cm]{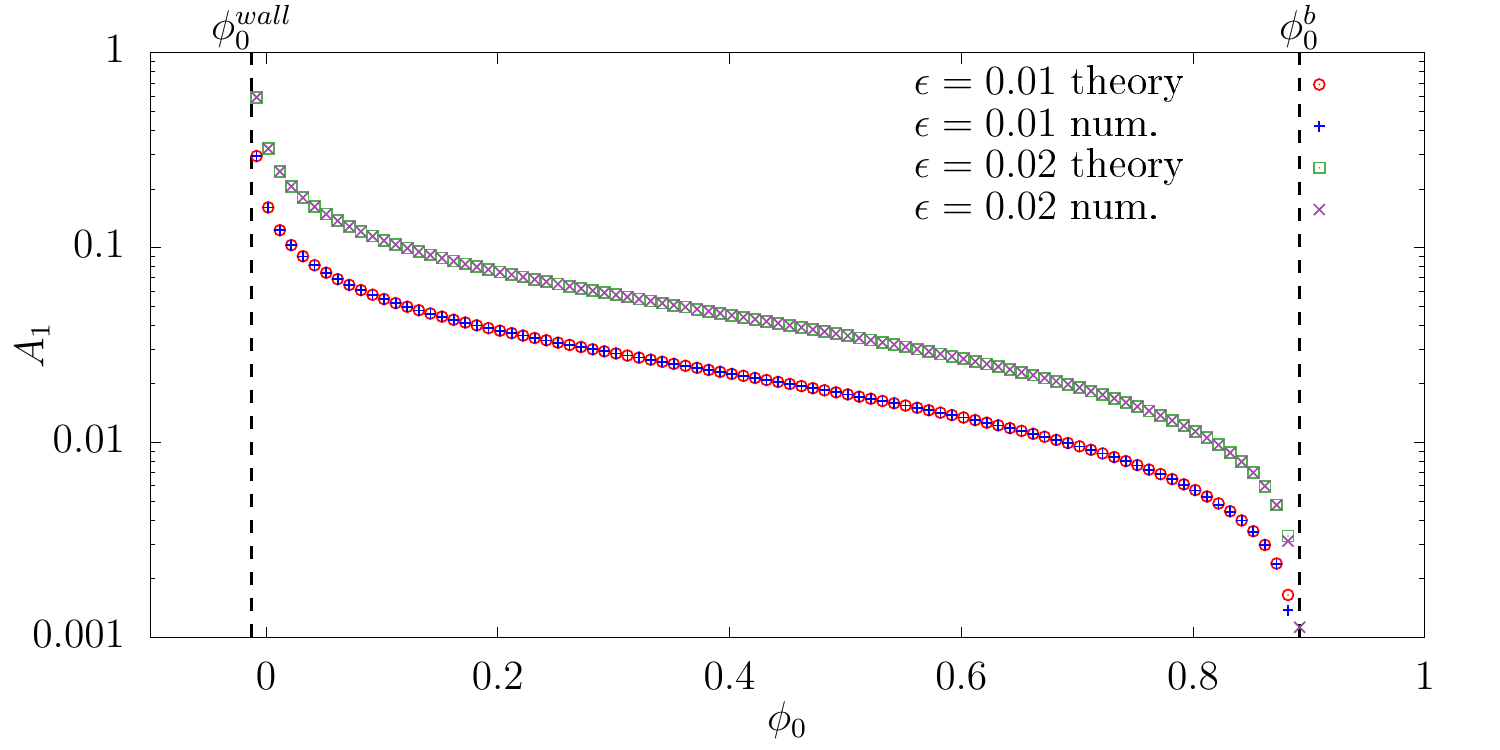}
\caption{Critical amplitude of the excited mode vs. the position of the stationary solution for $\epsilon=0.01$ and $0.02$.}
\label{saddle}
\end{figure}

To test the accuracy of the perturbative analysis, we calculated the amplitude $A_{1,{\rm c}}$ obtained
from eq. (\ref{A1}) as a function of $\phi_0$.
Note that in order to calculate the integrals, first we had to find the
eigenmodes $\eta_0$, $\eta_1$ and the appropriate eigenvalues $\omega_i^2$
numerically, all of which depend on $\phi_0$ and $\epsilon$. We solved
the eigenvalue problem using a shooting method, integrating the linearized
equation and matching with the exponential tail. Next, we calculated the
integrals approximating the eigenfuctions using the Hermite interpolation
method.

The eigenfunctions for a given position $\phi_0$ on the moduli space
were also used to prepare initial conditions for the evolution of the
full Euler-Lagrange equation. For a small amplitude of the excitation, the solitons attract
each other and collide. Too large amplitudes lead to repulsion. Using the
shooting method, we were able to find the amplitude which allows for the stationary (saddle point) 
solution, which  remains at the same position for a long time (of the order of
$t=200$--$1000$ depending on $\epsilon$ and $\phi_0$).

In Fig. \ref{saddle}, we compare the theoretical prediction (\ref{A1}) with the full numerical computation for $\epsilon= 0.01$ and $0.02$. 
Concretely, in the full numerical calculation we place an initially static SAS pair (a configuration $\phi(x,\phi_0)$) plus the excited mode at a {\em finite} distance ($\phi_0 <1$) and vary the mode parameter $A_1$ until we find a stationary solution (no force between the kink and antikink).   
We plot the amplitudes providing different stationary solutions and their related positions $\phi_0$, which, in principle, would correspond to different initial velocities $v$ for the full dynamical evolution from an initially widely separated SAS pair. The agreement is striking. In particular, for large $A_{1,{\rm c}}$ we approach the bottom of the thick wall (the original thin SW).

In addition, there is a simple scaling relation between the $A_{1,{\rm c}}$ vs $\phi_0$ curves for different $\epsilon$. Indeed, after multiplication by $\epsilon$ the curves coincide. This follows from eq. (\ref{A1}), where the first non-zero contribution in the nominator is of the order $\epsilon^2$ (the order $\epsilon^0$ vanishes due to the SD property of $\Phi$) while the denominator has a nontrivial part of the order $\epsilon^0$. This linear scaling may receive some corrections for larger $\epsilon$. 

We underline that in the weak static force regime the effective description does not collapse at the point where the stationary solution is found, because all relevant mode functions (i.e., $\eta_0$ and $\eta_1$)  remain normalizable at this point, and the projections on these modes are well-defined. Therefore, we have been able to compute the critical amplitude, which is, on the other hand, currently not accessible in the self-dual (no static force) limit. 
\section{Summary}
The main result of this paper is the observation that, once we allow for a non-zero static intersoliton force, which is a generic feature of all realistic soliton-antisoliton processes, thin spectral walls transmute into thick spectral walls with a certain stiffness. This means that, if we move from a BPS to a non-BPS SAS process, instead of a sharp selective barrier, whose position in principle does not change with the velocity of the scattered solitons, we find a finite region for the possible position of the barrier. Its (finite) stiffness implies that for larger velocities the barrier is located closer to the original thin SW. Further, the stiffness of the thick SW increases as we increase $\epsilon$, i.e., move away from the self-dual regime. For $\epsilon \rightarrow 0$ the thick SW basically disappears and we see only its bottom, i.e., the thin SW. As $\epsilon$ grows, the thick SW behaves more resistant, that is, for a given velocity the barrier (position of the stationary trajectory) is located further away from the original 
thin SW. Hence, if $\epsilon$ increases, the kink and the antikink feel the barrier sooner, while they are still further apart and their identity is more pronounced. This we interpreted as
the increasing stiffness of the thin SW when $\epsilon$ grows.

On the other hand, the size or extension of the thick SW does not change with $\epsilon$. Its bottom is always the thin SW (which is probably a model independent feature), while it ends at the position of the local maximum of the effective potential $\phi_0^b$. This latter position is certainly a particular property of our BPS breaking set-up (our way of introducing the intersoliton static force). One can imagine a situation where the thick SW extends to $\phi_0 \rightarrow 1$, if no maximum exists. 

The appearance of the thick SW can be viewed as an enhancement of the original thin SW phenomenon, which is a welcome effect, as most of the solitonic interactions (especially SAS scatterings) are strongly non-BPS processes. Furthermore, we also observe a weakening of the vacuum wall due to the non-BPS part of the model (the effective potential). As a consequence, the barrier is no longer distorted by the vicinity of the vacuum wall (see the thin SW in the self-dual limit) and now it is much better visible. 

The existence of the thin and thick spectral walls is based on different mechanisms. While the former is due to the mode transition to the continuum spectrum, the latter is based on the appearance of a stationary saddle point solution resulting from a compensation of the acceleration of the solitons (as dictated by the effective potential on the unstable manifold) by the sufficiently excited mode. Thus, now this barrier happens before the mode crosses the mass threshold. This is a nice feature, as it leads to a well defined effective theory, allowing for a quantitative description of the effect, see eq. (\ref{A1}). 

It is worth underlining that the BPS breaking (appearance of the intersolitonic forces) acts differently on thin spectral walls and vacuum walls. In the BPS limit, the vacuum wall is the leading factor of the dynamics beyond the geodesic approximation. It concerns all perturbations and can strongly affect thin spectral walls located in too close neighborhoods. When the BPS property is broken, and solitons interact via a static force, a thin spectral wall expands to a thick spectral wall which is much more pronounced. This especially concerns thin spectral walls which were hidden in the shadow of the vacuum wall. Furthermore, the vacuum wall is very quickly removed. So, in practice, it disappears from non-self-dual processes. Hence we can say that, in contrast to thin spectral walls, the VW seems to be a pure BPS effect. 

Although we considered a particular model, with a particular mode excited, it must be underlined that the results are much more general. The existence of a static force between a soliton and an antisoliton is a generic feature shared by any realistic model supporting topological solitons. Furthermore, thick spectral walls will exists in near-BPS SAS processes in any (1+1) dimensional theory, also for a model which supports (anti)kinks without oscillating modes. This is due to the fact that the initial configuration, i.e., an infinitely separated pair of kink and antikink, has two zero modes. Then, in the self-dual background field limit, only one zero mode survives, namely the symmetric superposition of the zero modes of the asymptotical states. The asymmetric superposition becomes a massive mode which necessarily crosses into the continuum during the collision. Hence the corresponding thin spectral wall shows up. In a weakly broken BPS process, this thin spectral wall will again be enhanced to a thick spectral wall. As a result, we expect that thick spectral walls will exist in various processes in many (1+1) dimensional field theories \cite{izquierdo}-\cite{bazeia}. Especially, they should be relevant for the recently discussed scattering of wobbling kinks \cite{izq}

Looking from a more general perspective, we extended the self-dual background field framework, which previously allowed to understand solitonic processes in given models in the limit of no static force, to a case where the BPS property is weakly broken. This means that a weak static intersoliton force appears. This regime is still mathematically well-defined and, as a consequence, allows for a rigorous mathematical analysis and a good understanding of the dynamics. 

In the final step, the limit $\epsilon \rightarrow 1$, which reproduces the original theory we would like to investigate, should be considered. That could, for example, prove the existence of the spectral walls in the original model. We will investigate this problem in a forthcoming paper. 

\section*{Acknowledgments}
 The Authors acknowledge financial support from the Ministry of Education, Culture, and Sports, Spain (Grant No. FPA2017-83814-P), the Xunta de Galicia (Grant No. INCITE09.296.035PR and Conselleria de Educacion), the Spanish Consolider-Ingenio 2010 Programme CPAN (CSD2007-00042), Maria de Maetzu Unit of Excellence MDM-2016-0692, and FEDER. KO acknowlegdes  support from the National Science Centre, Poland (Grant MINIATURA 3 No. 2019/03/X/ST2/01690). KO, TR and AW were supported by the Polish National Science Centre, 
grant NCN 2019/35/B/ST2/00059. Further, we thank F. Simas for carefully reading the manuscript and for spotting a sign error in Eq. (15). 

\renewcommand{\theequation}{A.\arabic{equation}}
\setcounter{equation}{0} 
\section*{Appendix A: The BPS solution as the weighted superposition of kink and antikink.} 
The BPS kink-antikink solution (\ref{sol-sd-a}) can be written as a weighted superposition of the kink and antikink of the original $\phi^4 $ theory located at $\mp c$ respectively \cite{MORW}. Indeed,
\be
\phi(x;a)=\frac{a-\cosh^2x}{a+\cosh^2 x} = \tanh c \left( \tanh(x+c) - \tanh(x-c) \right) -1 \label{decomp}
\ee
where $a=\sinh^2(c)$. Similarly, $\phi_0=(\sinh^2 c -1)/(\sinh^2 c +1)$. This decomposition holds for any $c \in \mathbb{R}$ and therefore for $a \ge 0$ or $\phi_0 \ge -1$. Hence, for these values the moduli space coordinate can uniquely identify the position of the solitons. At $c=0$ where $c=0$ and $\phi_0=-1$ the constituent kink and antikink are on top of each other. Interestingly, the decomposition (\ref{decomp}) is valid also for $a \in (-1, 0)$ (and $\phi_0 \in (-\infty, -1)$. However, it requires an imaginary $c=i\tilde{c}$, where $\tilde{c} \in (0, \pi/2)$. Now, $a=- \sin^2 \tilde{c}$. A possible interpretation is that now the solitons completely lose their identity. In any case, neither $c$ nor $\tilde c$ are good coordinates on the full moduli space, which is related to the so-called "null vector problem" \cite{MORW}. 

\renewcommand{\theequation}{B.\arabic{equation}}
\setcounter{equation}{0} 
\section*{Appendix B: Derivation of the position of the reflection point $\phi_0^*$} 
We begin with the energy conservation for the flow on the unstable manifold
\begin{equation}
\frac{1}{2} M(\phi_0^{in}) (\dot{\phi}_0^{in})^2 + \epsilon^2 V(\phi_0^{in}) = \epsilon^2 V(\phi_0^*) \label{energy-cond}
\end{equation}
where $\phi_0^{in}$ and $\dot{\phi}_0^{in}$ is the initial position on the moduli and its velocity. In the first step, this has to be related with the velocity $v$ of the initially very well separated kink-antikink state, i.e., for $\phi_0^{in} \to 1$. Comparing the exact self-dual solution (\ref{sol-sd}) with a widely separated $\phi^4$ kink-antikink pair (\ref{init-unpert}) we get
\begin{equation}
 \frac{(1+\phi_0) - (1-\phi_0)\cosh^2 x}{(1+\phi_0) +(1-\phi_0)\cosh^2 x} \simeq -\tanh (\gamma(x-x_0+vt))
+\tanh (\gamma(x+x_0-vt))-1. \label{initial}
\end{equation}
The two expressions at the left and right of the symbol $\simeq$ are never exactly equal, but they get arbitrarily close for a sufficiently separated SAS pair.
Further,  $\pm x_0$ are the initial positions (at $t=0$) of the kink and antikink, respectively. For well separated solitons, these initial positions are defined by the zeros of the above expression at $t=0$, which allows us to relate the initial value $\phi_0^{in}$ with the initial position $x_0$. Indeed, setting the left-hand-side equal to 0 for $x=x_0$ we find
\begin{equation} \label{rel-phi_in-x_0}
 \cosh^2 x_0 = \frac{1+\phi_0^{in}}{1-\phi_0^{in}}\simeq  \frac{2}{1-\phi_0^{in}}
 \end{equation}
where the limit $\phi_0^{in} \to 1$ in the nominator has been explicitly performed. Next, we differentiate eq. (\ref{initial}) w.r.t. time. This gives
 \begin{eqnarray}
\frac{\dot{\phi_0}}{\left((1+\phi_0) +(1-\phi_0)\cosh^2 x\right)^2} \left( (1+\cosh^2x)((1+\phi_0) +(1-\phi_0)\cosh^2 x)) - \right. \nonumber \\
\left. - (1-\cosh^2x)((1+\phi_0) -(1-\phi_0)\cosh^2 x))  \right) \simeq \nonumber \\
\simeq -\gamma v \left(\frac{1}{\cosh^2(\gamma(x+x_0-vt))} + \frac{1}{\cosh^2(\gamma(x-x_0+vt))}  \right) .
\end{eqnarray} 
Now we set $t=0$, $ x= x_0$ and use Eq. (\ref{rel-phi_in-x_0}) to find
\begin{equation}
\dot{\phi}_0^{in}\simeq -2 v\gamma (1-\phi_0^{in}). \label{init-phi-dot}
\end{equation}
Next, we observe that the moduli space metric (\ref{mod-metric}) for $\phi_0^{in} \to 1$ diverges as
\begin{equation}
M(\phi_0^{in})\simeq \frac{2}{3} \frac{1}{(1-\phi_0^{in})^2}. \label{init-metric}
\end{equation}
Finally, the effective potential (\ref{eff_pot}) for $\phi_0^{in} \to 1$ tends to a finite value
\begin{equation}
V(\phi_0^{in})\simeq \frac{8}{3}. \label{init-pot}
\end{equation}
Inserting the results (\ref{init-phi-dot}), (\ref{init-metric}), (\ref{init-pot}) into the energy conservation law (\ref{energy-cond}) we get 
\begin{equation}
\frac{4}{3} v^2\gamma^2 + \frac{8}{3}\epsilon^2=\epsilon^2 V(\phi_0^*),
\end{equation}
which after a trivial manipulation gives (\ref{core}).

\end{document}